\documentclass{article}
\usepackage[utf8]{inputenc}
\usepackage{rotating}

\usepackage{enumerate}
\usepackage{mathtools}
\mathtoolsset{showonlyrefs=true}
\usepackage{todonotes}

\usepackage{url}
\usepackage{amsfonts}
\usepackage{amsmath,amssymb,color}
\usepackage{graphicx}
\usepackage{epstopdf}
\usepackage{multirow}
\usepackage{bm}
\usepackage{natbib}

\usepackage{verbatim}
\usepackage{float}
\usepackage[toc,page]{appendix}
\usepackage{setspace}
\usepackage{setspace}
\usepackage{adjustbox}
\usepackage{rotating}
\usepackage{subcaption}
\usepackage{amsthm}
\usepackage{booktabs,dcolumn,caption}

\usepackage{color}
\usepackage{ulem}

\usepackage{tikz}
\usepackage{multirow}
\usepackage{graphicx}
\usepackage{adjustbox}
\usepackage{float}
\usepackage{footnote}
\usepackage{array}
\usepackage{longtable}
\usepackage{threeparttablex}
\usepackage{booktabs,dcolumn,caption}
\usepackage{amsmath,amssymb,color}
\usepackage{fnbreak}
\usetikzlibrary{arrows.meta,shapes.multipart,positioning}
\usepackage{mathtools}

\usepackage{mathabx}
\usepackage{array}
\usepackage{longtable}
\usepackage{threeparttablex}

\usepackage{chngcntr}
\usepackage{pdflscape}

\usepackage{diagbox} 
\usepackage{longtable}

\usepackage{tikz} 
\usepackage{algpseudocode,algorithm,algorithmicx}

\let\hat\widehat
\let\tilde\widetilde

\newcolumntype{L}[1]{>{\raggedright\let\newline\\\arraybackslash\hspace{0pt}}p{#1}}
\newcolumntype{C}[1]{>{\centering\let\newline\\\arraybackslash\hspace{0pt}}p{#1}}
\newcolumntype{R}[1]{>{\raggedleft\let\newline\\\arraybackslash\hspace{0pt}}p{#1}}

\newcommand{\yc}[1]{{{\color{black}  #1}}}

\newcommand{\zx}[1]{{{\color{black}  #1}}}
\newcommand{\jw}[1]{{{\color{black}  #1}}}

\newcommand{\argmin}{\operatornamewithlimits{arg\,min}}

\newtheorem{theorem}{Theorem}

\newtheorem{condition}{Condition}

\newtheorem{remark}{Remark}

\title{Exact Exploratory Bi-factor Analysis: A Constraint-based Optimisation Approach}
\author{Jiawei Qiao, Yunxiao Chen and Zhiliang Ying}
\date{}
\begin{document}

\maketitle

\begin{abstract}
Bi-factor analysis is a form of confirmatory factor analysis 
widely used in psychological and educational measurement. The use of a bi-factor model requires the specification of an explicit bi-factor structure on the relationship between the observed variables and the group factors. In practice, the bi-factor structure is sometimes unknown, in which case an exploratory form of bi-factor analysis is needed to find the bi-factor structure. Unfortunately, there are few methods for exploratory bi-factor analysis, with the exception of a rotation-based method proposed in \cite{jennrich2011exploratory,jennrich2012exploratory}.
However, the rotation-based method only finds approximate bi-factor structures, as it does not yield an exact bi-factor loading structure, even after applying hard thresholding. 
In this paper, we propose a constraint-based optimisation method that learns an exact bi-factor loading structure from data, overcoming the issue with the rotation-based method. The key to the proposed method is a mathematical characterisation of the bi-factor loading structure as a set of equality constraints, which allows us to formulate the exploratory bi-factor analysis problem as a constrained optimisation problem in a continuous domain and solve the optimisation problem with 
an augmented Lagrangian method. The power of the proposed method is shown via simulation studies and a real data example. Extending the proposed method to exploratory hierarchical factor analysis is also discussed. Code implementing the proposed method is open-source and publicly available at https://anonymous.4open.science/r/Bifactor-ALM-method-757D.

\bigskip
\noindent
Keywords: Bi-factor model, augmented Lagrangian method, exploratory bi-factor analysis, hierarchical factor model
\end{abstract}

\section{Introduction}\label{sec:intro}

The bi-factor model was originally proposed by \cite{holzinger1937bi} for linear factor analysis and further extended by \cite{gibbons1992full}, \cite{gibbons2007full}, \cite{cai2011generalized}, among others, to nonlinear factor analysis settings to account for dichotomous, ordinal, and nominal data. These models assume that the observed variables can be accounted for by $(G +1)$ factors, with a general factor, onto which all items load directly,  and $G$ group factors that each is associated with a subset of variables.
Such a specification leads to good interpretations in many real-world applications. These models have received wide applications in psychological and educational measurement; see e.g., \cite{bradlow1999bayesian,cai2016item,chen2012modeling,demars2006application,demars2012confirming,gibbons2009psychometric,gignac2013bifactor,jeon2013modeling,reise2007role,rijmen2010formal}.
However, we note that all these applications of bi-factor analysis are confirmatory in the sense that one needs to pre-specify the number of group factors and the relationship between the observed variables and the group factors. Such prior knowledge may not always be available.  In that case, an exploratory form of bi-factor analysis is needed. 

Exploratory bi-factor analysis can be seen as a special case of exploratory factor analysis, which dates back to the seminal work of \cite{Thurstone1947FABook} concerning finding a ``simple structure" of loadings. Various rotation methods have been proposed for exploratory factor analysis. A short list of relevant works includes \cite{kaiser1958varimax}, \cite{mccammon1966principal}, \cite{jennrich1966rotation}, \cite{mckeon1968rotation}, \cite{crawford1970general}, \cite{yates1987multivariate}, \cite{jennrich2006rotation}, \cite{jennrich2004rotation}, \cite{jennrich2011exploratory,jennrich2012exploratory}, and \cite{liu2022rotation}. We refer the readers to \cite{Browne2001MBR} for a review of rotation methods for exploratory factor analysis.

However, standard exploratory factor analysis methods do not apply to the bi-factor analysis setting, and few methods have been developed for exploratory bi-factor analysis. An exception is the seminal work of \cite{jennrich2011exploratory,jennrich2012exploratory}, who proposed a rotation-based method for exploratory bi-factor analysis with orthogonal or oblique factors. However, their approach has some limitations. First, as a common issue with rotation-based methods, their method does not yield many zero loadings, and thus, the resulting loading structure does not have an exact bi-factor structure. Although a post-hoc thresholding procedure (i.e., treating loadings with an absolute value below a threshold as zero) can be applied to obtain a cleaner loading pattern, it does not work well when some variables show relatively large loadings on more than one group factor after the rotation.  In fact, one cannot always find a threshold that yields an exact bi-factor structure that each variable loads on one and only one group factor. 
Second, as noted in \cite{jennrich2012exploratory}, their method fails completely in the best case where there is a rotation of an initial loading matrix that has an exact bi-factor structure. This failure is due to that their rotation method cannot incorporate the zero constraints on the correlations between the general factor and the group factors.

This paper proposes a constrained optimisation method for exploratory bi-factor analysis, which overcomes the issues with the rotation-based method. The contribution is four-fold. First, we provide a mathematical characterisation of the
bi-factor loading structure as a set of nonlinear equality constraints, which allows us to formulate the exploratory bi-factor analysis problem as a constrained optimisation problem. In other words, it turns a discrete model selection problem into a continuous optimisation problem, which reduces the computational demand in some sense. 
It is shown that in the aforementioned best case where the rotation method fails, 
the global solutions to the optimisation can perfectly recover the true bi-factor structure. Second, we propose an augmented  Lagrangian method \citep[ALM,][]{bertsekas2014constrained} for solving this optimisation problem, which is a standard numerical optimisation method for solving constrained optimisation with robust empirical performance and good theoretical properties. Third, we combine the proposed method with the Bayesian information criterion \citep[BIC,][]{schwarz1978estimating} for selecting the number of group factors. Compared with existing exploratory factor analysis methods for determining the number of factors, our method is tailored to the bi-factor model structure and, thus, tends to be statistically more efficient when the data is indeed generated by a bi-factor model.  
Finally, we demonstrate that the proposed method can be extended to learning the loading structure of hierarchical factor models \citep{schmid1957development,yung1999relationship}, a higher-order extension of the bi-factor model that has received wide applications \citep[see, e.g.,][and references therein]{brunner2012tutorial}. The bi-factor model can be viewed as a special hierarchical factor model with a two-layer factor structure, with the general factor in one layer and the group factors in the other. 
Similar to exploratory bi-factor analysis, the proposed method yields exact hierarchical factor loading structures without a need for post-hoc treatments.

The rest of the paper is organised as follows. In Section~\ref{sec:explore bifactor}, we formulate the exploratory bi-factor analysis problem as a constrained optimisation problem and propose an ALM for solving it. We also propose a BIC-based procedure for selecting the number of group factors. Simulation studies and a real data example are presented in Sections \ref{sec:sim} and \ref{sec:real data}, respectively, to evaluate the performance of the proposed method. We conclude with discussions in Section~\ref{sec:diss}.
The appendix in the supplementary material includes additional details about the computation, the simulation studies and the real data example, an extension of the proposed method to exploratory hierarchical factor analysis, and proof of the theoretical results.

\section{Exploratory Bi-factor Analysis by Constrained Optimisation}\label{sec:explore bifactor}

\subsection{Bi-factor Model Structure and a Constrained Optimisation Formulation}\label{subsec:opti}

For the ease of exposition and simplification of the notation, we focus on the linear bi-factor model, while noting that the constraints that we derive for the bi-factor loading matrix below can be combined with the likelihood function of other bi-factor models \citep[e.g.,][]{gibbons1992full,gibbons2007full,cai2011generalized} for their exploratory analysis. 
We focus on the extended bi-factor model, also known as the oblique bi-factor model, as considered in \cite{jennrich2012exploratory} and \cite{fang2021identifiability}. This model is more general than the standard bi-factor model, in the sense that the latter assumes all the factors to be uncorrelated while the former allows the group factors to be correlated.  
As established in \cite{fang2021identifiability}, this extended bi-factor model is identifiable under mild conditions. We also point out that the proposed method can be easily adapted to the standard bi-factor model.

Consider a dataset with $N$ observation units from a certain population
and $J$ observed variables. {The extended bi-factor model assumes that there exists a general factor and $G$ group factors. \yc{The group factors are loaded by independent clusters of variables, 
where each variable belongs to only one cluster.} The model further assumes that the population covariance matrix of the observed variables can be decomposed as 
$$\Sigma = \Lambda \Phi \Lambda^\top + \Psi,$$
where $\Lambda = (\lambda_{jg})_{J \times (G+1)}$ is the loading matrix, $\Phi = (\phi_{gg'})_{(G+1)\times (G+1)}$ is the correlation matrix of the factors, which is assumed to be strictly positive definite, and $\Psi$ is a $J \times J$ diagonal matrix with diagonal entries $\psi_1$, ..., $\psi_J$. 
Let $\mathcal B_g \subset \{1, ..., J\}$ denote the cluster of variables loading on the $g$th group factor. Then the bi-factor model assumption implies that $\mathcal B_g \cap \mathcal B_{g'} = \emptyset$, $g\neq g'$, and $\cup_{g=1}^G \mathcal B_g = \{1, ..., J\}$. 
The following zero constraints on the loading matrix $\Lambda$ hold: 
$$\lambda_{j,g+1} = 0 \mbox{~for all~} j \notin \mathcal B_g.$$
In addition, the correlation matrix $\Phi$ satisfies 
$\phi_{1k} = 0$ for all $k\neq 1$, meaning that all the group factors are uncorrelated with the general factor. This constraint on $\Phi$ is necessary to ensure that the extended bi-factor model is identifiable \citep{fang2021identifiability}, as otherwise, there will be a rotational indeterminacy issue. 

Now suppose that the number of group factors $G$ is known, while the clusters $\mathcal B_g, g = 1, ..., G$, are 
unknown. Section~\ref{subsec:selection} considers the selection of $G$ when it is unknown. The bi-factor structure means that for each $j$, there is at most one non-zero element in $(\lambda_{j,2}, ..., \lambda_{j,G+1})^\top$. Consequently, the loading matrix $\Lambda$ should satisfy the following $J(G-1)G/2$ constraints: 

\begin{equation}\label{eq:constraint}
    \lambda_{jk}\lambda_{jk'} = 0, \mbox{~for all~} k,k' = 2, ..., G+1, k\neq k', j = 1, ..., J.
\end{equation}

Therefore, the exploratory bi-factor analysis problem can be translated into the following constrained optimisation problem
\begin{equation}\label{eq:optimisation}
\begin{aligned}
\min_{\Lambda, \Phi, \Psi}~~~ &  l( \Lambda \Phi \Lambda^\top + \Psi; S) \\
s.t. ~~~ &     \lambda_{jk}\lambda_{jk'} = 0, \mbox{~for all~} k,k' = 2, ..., G+1, k\neq k', j = 1, ..., J,\\
& \phi_{1k} = 0, k = 2, ..., G+1,   \Phi \mbox{~is correlation matrix},\\
&\mbox{and~} \Psi \mbox{~is a diagonal matrix},
\end{aligned}
\end{equation}
where $l$ is a loss function and $S$ is the sample covariance matrix of observed data. We focus on the case when $l$ is the fit function based on the normal likelihood 
\begin{equation}
l( \Lambda \Phi \Lambda^\top + \Psi;S) = N\big(\log(\text{det}(\Lambda \Phi \Lambda^\top + \Psi)) + \textnormal{tr}(S (\Lambda \Phi \Lambda^\top + \Psi)^{-1}) -\log(\text{det}(S)) - J\big),  
\end{equation} 
while noting that this loss function can be replaced by other loss functions for factor analysis \citep[see, e.g.,][]{chen2023estimation}, including the Frobeneious norm of $ \Lambda \Phi \Lambda^\top + \Psi -S$ that is used in the least square estimator for factor analysis.  
\yc{We can also replace the sample covariance matrix in \eqref{eq:optimisation} with the sample correlation matrix, which is equivalent to performing exploratory bi-factor analysis based on variance-standardised variables.}

The following theorem shows that the proposed method can perfectly recover the true bi-factor loading structure in the best case when {$S = \Sigma^{*}$, where $\Sigma^{*}$ is the true covariance matrix of data.} Note that the rotation method proposed in \cite{jennrich2012exploratory} completely fails in this case. 
Before giving the statement of the theorem, we introduce some additional notations. For any matrix $A=(a_{i,j})_{i=1,\ldots,m,j=1,\ldots,n}$, $\mathcal{S}_1\subset\{1,\ldots,m\}$ and $\mathcal{S}_{2}\subset \{1,\ldots,n\}$, let $A[\mathcal{S}_1,\mathcal{S}_2] = (a_{i,j})_{i\in\mathcal{S}_1, j\in\mathcal{S}_2}$ {be the submatrix of $A$ consisting of elements that lie in rows belonging to set $\mathcal{S}_1$ and columns belonging to set $\mathcal{S}_2$. \jw{For example, consider a matrix $A$ with more than two rows and three columns. With index sets $\mathcal S_1 = \{1,2\}$ and $\mathcal S_2 = \{1,3\}$, the 
submatrix  $A[\mathcal{S}_1,\mathcal{S}_2]$ is a two-by-two matrix, taking the form 
$$A[\mathcal{S}_1,\mathcal{S}_2] = A[\{1,2\},\{1,3\}] = \left(\begin{array}{cc}
    a_{11} & a_{13}  \\
   a_{21}  & a_{23}
\end{array}\right).$$} For any set $\mathcal{S}$, let $\vert\mathcal{S}\vert$ be the cardinality of $\mathcal{S}$. 

Let $\{\mathcal{B}_{g}^{*}, g=1, \ldots, G\}$ be the true non-overlapping clusters of the $J$ variables, satisfying for each $j\in \mathcal{B}_{g}^*$, $\lambda^{*}_{j,g+1} \neq 0$, $g = 1,\ldots,G$. Further let $\mathcal{H}^{*} = \{g | \Lambda^{*}[B^{*}_{g},\{1,1+g\}] \mbox{~has column rank~} 2\}$ be the set of group factors for which the group loadings are linearly independent of the corresponding common loadings. 
Let $\mathcal{D}$ be the set of diagonal matrix with its diagonal entries taking values either $1$ or $-1$, and $\mathcal{P}$ be the set of permutation matrix $P$ such that each row and column of $P$ has exactly one nonzero entry of value 1 and $P_{11} = 1$. \yc{Each matrix in $\mathcal {D}$ corresponds to a simultaneous sign flip of certain factors and the corresponding loading parameters. Each matrix in $\mathcal{P}$ corresponds to a swapping of certain columns in the loading matrix associated with the group factors or, equivalently, a relabelling of the group factors. They are introduced to account for the sign-indeterminacy of the $G+1$ factors and the label indeterminacy of the group factors, respectively.} See Theorem~\ref{thm:id} and Remark~\ref{rmk:indet} below for more explanations.

{Let $\Lambda^*$, $\Phi^*$ and  $\Psi^*$ be the true values of the corresponding parameter matrices.} The following conditions are sufficient for the identifiability of the bi-factor structure and its parameters. 
\begin{condition}\label{cond:seperate}
Given $S = \Sigma^* = \Lambda^* \Phi^* (\Lambda^*)^\top + \Psi^*$. Suppose that there exists another pair of parameters $\Lambda, \Phi, \Psi$ satisfy the bi-factor structure constraints, we have $\Lambda^* \Phi^* (\Lambda^*)^\top = \Lambda \Phi (\Lambda)^\top$ and $\Psi^* = \Psi$.
\end{condition}

\begin{condition}\label{cond:E3Sfang}
$\vert \mathcal{H}^{*}\vert \geq 2$. In addition, there exists $g_1 \in \mathcal{H}^{*}$ such that 
$\vert \mathcal{B}_{g_1}^{*}\vert\geq3$ and any 2 rows of $\Lambda^{*}[\mathcal{B}^{*}_{g_1},\{1, 1+g_1\}]$ are linearly independent.
\end{condition}

\begin{theorem}\label{thm:id}
Suppose that Conditions~\ref{cond:seperate} and \ref{cond:E3Sfang} hold. For any parameters $\Lambda, \Phi, \Psi$ that satisfy $S = \Sigma^{*} = \Lambda \Phi (\Lambda)^\top + \Psi$, there exist a diagonal sign-flip matrix $D\in \mathcal{D}$ and a permutation matrix $P\in\mathcal{P}$ such that $\Lambda = \Lambda^{*}PD$ and $\Phi = DP^{\top}\Phi^{*}PD $. 
\end{theorem}
The proof of Theorem~\ref{thm:id} is given in Appendix~\ref{appen:Proofthm}.

\begin{remark}\label{rmk:indet}
We note that without additional information, the best we can achieve is to recover $\Lambda^*$ and $\Phi^*$ up to $\Lambda = \Lambda^{*}PD$ and $\Phi = DP^{\top}\Phi^{*}PD$, where 
the permutation matrix $P$ and sign-flip matrix $D$ are necessary to account for the label and sign indeterminacies of the factor model. In that case, $\Lambda^* \Phi^* (\Lambda^*)^\top = \Lambda \Phi (\Lambda)^\top$, and thus, the model implied covariance matrix is the same. \yc{A similar indeterminacy issue also appears in exploratory factor analysis; see, e.g., Remark 1 in \cite{liu2022rotation}.}

\end{remark}

\begin{remark}
Condition~\ref{cond:seperate} ensures the separation between low rank matrix $\Lambda^* \Phi^* (\Lambda^*)^\top$ and diagonal matrix $\Psi^*$. A sufficient condition for Condition~\ref{cond:seperate} that can be easily checked in practice is given in Condition~\ref{cond:altersep} below, {which requires that each group has at least three non-zero group loadings and there exist at least three groups whose group loadings are linearly independent of the corresponding common loadings.}
We refer to Theorem 5.1 in \cite{anderson1956statistical} and Theorem 2 in \cite{fang2021identifiability} for alternative sufficient conditions of Condition~\ref{cond:seperate}. 
\end{remark}

\begin{condition}\label{cond:altersep}
$ \vert\mathcal{B}^{*}_{g}\vert\geq 3$ for all $g=1,\ldots,G$ and $\vert\mathcal{H}^{*}\vert\geq3$. 
\end{condition}

\begin{remark}
\jw{Condition \ref{cond:E3Sfang} is similar to Condition E3S for Proposition 1 in \cite{fang2021identifiability}, where the latter is used to ensure the identifiability of parameters when the bi-factor structure is known.
It is a sufficient condition that ensures if there is a pair of $\Lambda$ and $\Phi$ that also satisfies the constraints of a bi-factor model and $\Lambda\Phi\Lambda^\top = \Lambda^*\Phi^*(\Lambda^*)^\top$,  then there must be a permutation matrix $P$ and sign-flip matrix $D$  such that   $\Lambda = \Lambda^{*}PD$ and $\Phi = DP^{\top}\Phi^{*}PD$. This condition first requires the existence of at least two group factors, for each of which the group loadings are linearly independent of the corresponding common loadings. It further requires that there exists a group factor $g_1$ among these group factors, such that (1) $g_1$ has at least three nonzero group loadings, and (2) any two-by-two submatrix of $\Lambda^*$, whose rows correspond to any two variables loading on $g_1$ and columns correspond to the common factor and the group factor $g_1$, is of full rank.}
We note that the requirement of Condition \ref{cond:E3Sfang} is very mild. In fact, the set of parameters not satisfying this condition has zero Lebesgue measure in the parameter space for bi-factor models {satisfying that there are at least two group factors $g_1$ and $g_2$ such that $\vert\mathcal{B}^{*}_{g_1}\vert \geq 3$ and $\vert\mathcal{B}^{*}_{g_2}\vert \geq 2$. }

\end{remark}

\subsection{Proposed ALM}\label{subsec:alm}

Following the previous discussion, we see that we can perform exploratory bi-factor analysis by solving the optimisation problem with some equality constraints and the constraint that $\Phi$ is a correlation matrix. {To deal with the constraints in $\Phi$, we consider a reparameterisation of $\Phi$ based on a Cholesky decomposition, where the explicit form of the reparametrisation is given in Appendix~\ref{appen:repara}. With slight abuse of notation, we reexpress the covariance matrix as $\Phi(\boldsymbol\gamma)$, where $\boldsymbol\gamma$ is a $G(G-1)/2$ dimensional unconstrained parameter vector.} In addition, we use $\boldsymbol\psi = (\psi_1, ..., \psi_{J})^\top$ to denote the vector of diagonal entries of $\Psi$ and reexpress the residual covariance matrix as $\Psi(\boldsymbol\psi)$. Thus, the optimisation problem \eqref{eq:optimisation} is now simplified as 
\begin{equation}\label{eq:optimisation2}
\begin{aligned}
\min_{\Lambda, \boldsymbol{\gamma}, \boldsymbol \psi}~~~ &  l( \Lambda \Phi(\boldsymbol \gamma) \Lambda^\top + \Psi(\boldsymbol{\psi}); S) \\
s.t. ~~~ &     \lambda_{jk}\lambda_{jk'} = 0, \mbox{~for all~} k,k' = 2, ..., G+1, k\neq k', j = 1, ..., J, 
\end{aligned}
\end{equation}
which is an equality-constrained optimisation problem. 

The standard approach for solving such a problem is the augmented Lagrangian method \citep[e.g.,][]{bertsekas2014constrained}. This method aims to find a solution to \eqref{eq:optimisation2} by solving a sequence of unconstrained optimisation problems. Let $t$ denote the $t$th unconstrained optimisation problem in the ALM. The corresponding objective function, also known as the augmented Lagrangian function,
takes the form 

\begin{equation}\label{eq:optimisation3}
\min_{\Lambda, \boldsymbol \gamma, \boldsymbol \psi}~~~  l( \Lambda \Phi(\boldsymbol \gamma) \Lambda^\top + \Psi(\boldsymbol{\psi}); S) + \left(\sum_{j=1}^J \sum_{k=2}^G \sum_{k'=k+1}^{G+1} \beta_{jkk'}^{(t-1)} \lambda_{jk}\lambda_{jk'} \right)+  \frac{1}{2}c^{(t-1)} \left(\sum_{j=1}^J \sum_{k=2}^G \sum_{k'=k+1}^{G+1} (\lambda_{jk}\lambda_{jk'})^2 \right),
\end{equation}
where  $c^{(t-1)} > 0$  and $\beta_{jkk'}^{(t-1)}$s are auxiliary coefficients of the ALM determined by the initial values when $t=1$ and the previous optimisation when $t\geq 2$. Details of the ALM are given in Algorithm~\ref{alg:main} below, in which the function $h$ returns the second-largest value of a vector.

\jw{The ALM can be seen as a penalty method for solving constrained optimization problems. It replaces a constrained optimization problem with a series of unconstrained problems. It adds a penalty term, i.e., the third term in \eqref{eq:optimisation3}, to the objective to enforce the constraints. The tuning parameter $c^{(t-1)}$ can be seen as the weight of the penalty term. 
In fact, as $c^{(t)}$ goes to infinity while $\beta_{jkk'}^{(t)}$s remain bounded, the solution has to converge to one satisfying the 
equality constraints in \eqref{eq:optimisation2}, as otherwise, the objective function value in \eqref{eq:optimisation3} will diverge to infinity. However, the ALM is not purely a penalty method in the sense that it also adds the second term in \eqref{eq:optimisation3} to mimic a Lagrange multiplier (see, e.g., Chapter 12, \citealp{nocedal1999numerical}), for which $\beta_{jkk'}^{(t-1)}$s are the weights. An advantage of the ALM is that, with the inclusion of the Lagrangian term (i.e., the second term), the method is guaranteed to converge to a local solution satisfying the equality constraints without requiring 
$c^{(t)}$ to go to infinity. This is important, as when $c^{(t)}$ is very large, the optimisation problem \eqref{eq:optimisation3} becomes ill-conditioned and thus hard to solve.}

\begin{algorithm}[ht!]
\caption{Augmented Lagrangian Method for Exact Exploratory Bi-factor Analysis  }
\label{alg:main}
\begin{algorithmic}[1]
\Require Initial value $\Lambda^{(0)}$, $\boldsymbol{\gamma}^{(0)}$ and $\boldsymbol{\psi}^{(0)}$, initial Lagrangian parameters $\beta_{j,k,k'}^{(0)}$ for $j=1,\ldots,J$, $k=2,\ldots,G$ and $k'=k+1,\ldots,G+1$, initial penalty coefficient  $c^{(0)}>0$, constants $c_{\theta} \in (0,1)$ and $c_{\sigma} > 1$, tolerances $\delta_1, \delta_2>0$.

\While{$t = 1,2,\ldots$}
\State Solve the following problem:
\begin{equation*}
\begin{aligned}
&\Lambda^{(t)}, \boldsymbol{\gamma}^{(t)}, \boldsymbol{\psi}^{(t)}\\
=&\arg\min_{\Lambda, \boldsymbol \gamma, \boldsymbol \psi}~~~  l( \Lambda \Phi(\boldsymbol \gamma) \Lambda^\top + \Psi(\boldsymbol{\psi}); S) + \left(\sum_{j=1}^J \sum_{k=2}^G \sum_{k'=k+1}^{G+1} \beta_{jkk'}^{(t-1)} \lambda_{jk}\lambda_{jk'} \right)+  \frac{1}{2}c^{(t-1)} \left(\sum_{j=1}^J \sum_{k=2}^G \sum_{k'=k+1}^{G+1} (\lambda_{jk}\lambda_{jk'})^2 \right).
\end{aligned}
\end{equation*}
\State Update $\beta_{jkk'}^{(t)}$ and $c^{(t)}$ according to equations \eqref{eq:beta rule} and \eqref{eq:c rule} 

\begin{equation}\label{eq:beta rule}
\begin{aligned}
\beta_{jkk'}^{(t)} = \beta_{jkk'}^{(t-1)} + c^{(t-1)}\lambda^{(t)}_{jk}\lambda^{(t)}_{jk'},
\end{aligned}
\end{equation}

and

\begin{equation}\label{eq:c rule}
\begin{aligned}
c^{(t)} = \left\{
    \begin{array}{l}
        c_{\sigma}c^{(t-1)} \mbox{~~if~~} \left(\sum_{j=1}^J \sum_{k=2}^G \sum_{k'=k+1}^{G+1} (\lambda_{jk}^{(t)}\lambda_{jk'}^{(t)})^2\right)^{1/2} > c_{\theta}\left(\sum_{j=1}^J \sum_{k=2}^G \sum_{k'=k+1}^{G+1} (\lambda_{jk}^{(t-1)}\lambda_{jk'}^{(t-1)})^2\right)^{1/2}; \\
        c^{(t-1)} \mbox{~otherwise}.
    \end{array}
\right.
\end{aligned}
\end{equation}

\If{$\left(\Vert \Lambda^{(t)} - \Lambda^{(t-1)} \Vert_{F}^{2} + \Vert \boldsymbol \gamma^{(t)} -  \boldsymbol \gamma^{(t-1)}  \Vert^{2} + \Vert \boldsymbol \psi^{(t)} -  \boldsymbol \psi^{(t-1)}  \Vert^{2}\right)^{1/2}/ \sqrt{J(G+2) + G(G-1)/2}<\delta_1,$\\
and $\max_{j \in \{1, ..., J\}} h(|\lambda_{j2}^{(t)}|, ..., |\lambda_{j,G+1}^{(t)}|)<\delta_2$} 
\State \textbf{Break}
\EndIf
\EndWhile
\Ensure $\Lambda^{(t)}, \boldsymbol{\gamma}^{(t)}, \boldsymbol{\psi}^{(t)}$.
\end{algorithmic}
\end{algorithm}

{The updating rule of $\beta^{(t)}_{jkk'}$ and $c^{(t)}$ follows equation (1) and (47) in Chapter 2.2 of \cite{bertsekas2014constrained}. The updating rule for $\beta_{jkk'}^{(t)}$ follows the first-order optimality condition based on optimisations \eqref{eq:optimisation2} and \eqref{eq:optimisation3}. 
The updating rule for $c^{(t)}$ ensures that it will become sufficiently large, which is necessary to guarantee the solution of the algorithm to converge to the feasible region defined by the zero constraints. On the other hand, it also prevents $c^{(t)}$ from growing too quickly with which the optimisation \eqref{eq:optimisation3} is ill-conditioned. As shown in Chapter 2.2 of \cite{bertsekas2014constrained}, as long as the sequences $\{\beta_{jkk'}^{(t)}\}$ remain bounded, the sequence $\{c^{(t)}\}$ remains bounded. We follow the recommended choices of  $c_{\theta} = 0.25$ and $c_{\sigma} = 10$ in \cite{bertsekas2014constrained}, while pointing out that the performance of Algorithm~\ref{alg:main} is quite robust against the choices of these tuning parameters; see Appendix~\ref{appen:sensitivity} for a sensitivity analysis. The convergence of Algorithm \ref{alg:main} is guaranteed by Proposition 2.7 of \cite{bertsekas2014constrained}.}

We remark on the stopping criterion in the implementation of Algorithm~\ref{alg:main}. We monitor the convergence of the algorithm based on two criteria: (1) the change in parameter values in two consecutive steps, measured by $$\left(\Vert \Lambda^{(t)} - \Lambda^{(t-1)} \Vert_{F}^{2} + \Vert \boldsymbol \gamma^{(t)} -  \boldsymbol \gamma^{(t-1)}  \Vert^{2} + \Vert \boldsymbol \psi^{(t)} -  \boldsymbol \psi^{(t-1)}  \Vert^{2}\right)^{1/2}/ \sqrt{J(G+2) + G(G-1)/2},$$
where $\Vert\cdot\Vert_F$ denotes the Frobeneious norm of a matrix and $\Vert\cdot \Vert$ denotes the standard Euclidian norm, and (2) the distance between the estimate and the space of bi-factor loading matrices measured by 
$$\max_{j \in \{1, ..., J\}} h(|\lambda_{j2}^{(t)}|, ..., |\lambda_{j,G+1}^{(t)}|).$$ 
We stop the algorithm when both criteria are below their pre-specified thresholds, $\delta_1$ and $\delta_2$, respectively.
\jw{The first criterion is a standard criterion for monitoring parameter convergence. This criterion being sufficiently small suggests the convergence of the algorithm. The second criterion is used to ensure that the solution is sufficiently close to the feasible set of optimisation defined by the equality constraints. This criterion being below $\delta_2$ means that for each variable $j$, there can only be one group loading whose absolute value is above the threshold $\delta_2$, and all the rest have absolute values below the threshold. 
Based on this, we can obtain an estimate of the bi-factor structure.} More specifically, let $T$ be the last iteration number. Then the estimated bi-factor model structure is given by 
$$\hat{\mathcal B}_g = \{j: |\lambda_{j,g+1}^{(T)}| > \delta_2\}.$$
By our choice of the stopping criterion, the resulting $\hat{\mathcal B}_g$, $g=1, ..., G$, gives a partition of all the variables, and thus, the bi-factor structure is satisfied. For simulation studies in Section~\ref{sec:sim}, we choose $\delta_1 = \delta_2 = 10^{-2}$. For real data analysis in Section~\ref{sec:real data}, we choose $\delta_1 = \delta_2 = 10^{-4}$ to get a more accurate and reliable result. 

The optimisation problem~\eqref{eq:optimisation3} is non-convex and can get stuck in a local minimum. Thus, we recommend running the proposed algorithm multiple times with random starting points and choosing the solution with the smallest objective function value. 
The algorithm can also suffer from slow convergence, especially when the penalty term becomes large. When the algorithm does not converge within $T_{\max}$ iterations, we suggest using the estimated parameters at the $T_{\max}$th iteration as the initial parameters and restarting the optimisation until a good proportion of them converge. In the simulation study in Section~\ref{sec:sim} below, the estimated parameters obtained using 50 random starting points are close to the global minimum in most cases in the simulation study. For the real data example in Section~\ref{sec:real data}, 200 random starting points are used to ensure a reliable result. We set {$T_{\max} = 100$} in all of our numerical studies. 

\subsection{Selecting the Number of Group Factors}\label{subsec:selection}

In Sections \ref{subsec:opti} and \ref{subsec:alm}, the number of group factors $G$ is treated as known. In practice, we can select its value based on the BIC \citep{schwarz1978estimating}. Let $l_G$ denote the minimum loss function value in \eqref{eq:optimisation} when the number of group factors is $G$. As $l_G$ differs from twice the negative log-likelihood of the bi-factor model with $G$ group factors by a constant, and the numbers of nonzero parameters in $\Lambda$ and $\Psi$ do not depend on $G$, it is not difficult to see that the BIC of the bi-factor model with $G$ group factors differs from
$l_G +  ({(G-1)G} \log(N))/2$ 
by a constant. Note that $(G-1)G/2$ is the number of free parameters in the correlation matrix $\Phi$. Thus, we write 
$$\text{BIC}_G = l_G +  ({(G-1)G} \log(N))/2.$$

In practice, we choose the number of group factors $G$ from a candidate set $\mathcal G$. For each value of $G\in \mathcal G$, we run the ALM described in Section~\ref{subsec:alm} to obtain the value of $l_G$. We then compute $\text{BIC}_G$ and choose $\hat G$ with the smallest BIC value, i.e., 
$$\hat{G} = \argmin_{G\in \mathcal G} ~\text{BIC}_G.$$

\section{Simulation Study}\label{sec:sim}

\subsection{Study I}\label{subsec:study1}
In this study, we compare the proposed method with the oblique bi-factor rotation in \cite{jennrich2012exploratory} regarding their performance in the estimation accuracy of parameters and the recovery of the bi-factor structure. \zx{We consider two different settings for the data generation mechanism: (1) the observed data are generated from an exact bi-factor model, and (2) the observed data are generated from an approximate bi-factor model, where the loading matrix is generated by adding small perturbations to an exact bi-factor loading matrix. } 

The oblique bi-factor rotation method first estimates the loading matrix $\hat{\Lambda}$ under the exploratory factor analysis setting by the optimisation problem
\begin{equation}\label{eq:efa}
\hat{\Lambda},\hat{\psi} = \argmin_{\Lambda\in\mathbb{R}^{J\times (G+1)},\psi\in\mathbb{R}^{J}} l(\Lambda\Lambda^{\top}+\Psi(\psi);S).
\end{equation}
We restrict $\lambda_{ij} = 0$ for $i = 2,\ldots,G$ and $j = i+1,\ldots,G+1$ to avoid the rotational indeterminacy of $\hat{\Lambda}$, as suggested in \cite{anderson1956statistical}. Then the rotated solutions $\hat{\Lambda}^{oblq}$ and $\hat{\Phi}^{oblq}$ are obtained by finding a rotation matrix that solves the optimisation problem for oblique bi-factor rotation \citep{jennrich2012exploratory}. The implementation in the R package GPArotation \citep{bernaards2005gradient} is used for solving this optimisation problem, which is based on a gradient projection algorithm. The optimisation problem for rotation is also nonconvex and thus may converge to local solutions. For a fair comparison, we also use 50 random starting points for the initial rotation matrix, which is the same as the number of random starting points that are used when running Algorithm~\ref{alg:main}.  

We first examine the accuracy in estimating the loading matrix. We calculate the mean squared error (MSE) for $\hat{\Lambda}$, after adjusting for the label and sign indeterminacy as considered in Theorem~\ref{thm:id} and further discussed in Remark~\ref{rmk:indet}. More specifically, let $\mathcal{P}$ and $\mathcal D$ be the sets of permutation and sign flip matrices, respectively, as defined in Theorem~\ref{thm:id}. We define the MSE for $\hat \Lambda$ as 
$$\text{MSE}_{\hat \Lambda} = \min_{P\in\mathcal{P}, D\in \mathcal D}\Vert \hat{\Lambda} - \Lambda^{*}PD\Vert_{F}^{2}/(J(1+G)).$$
\zx{Note that when data are generated from an approximate bi-factor model, $\Lambda^*$ does not have an exact bi-factor structure.}
This MSE is calculated for the loading matrix estimates from both methods. 

To compare the two methods in terms of their performance in recovering the bi-factor structure, we derive a sparse loading structure from the rotated solution by hard thresholding, a procedure also performed in \cite{jennrich2012exploratory} for examining structure recovery. 
We let 
$$\hat{\mathcal B}_{g}^{oblq} = \{j: |\lambda_{j,g+1}^{oblq}| > \delta\},$$
for $g = 1,\ldots,G$ and some hard thresholding parameter $\delta>0$. \zx{In the analysis below, we consider three choices of hard thresholding parameter $\delta\in\{0.1,0.2,0.3\}$.}
We note that $\hat{\mathcal B}_{g}^{oblq}$, $g = 1, ..., G$, may not yield an exact bi-factor structure as it is not guaranteed to return only one nonzero group loading parameter for each variable.

Let $\{\mathcal{B}_{g}^{*}, g=1, \ldots, G\}$ be the true non-overlapping clusters of the $J$ variables, and let $\{\mathcal{B}_{g}, g = 1,\ldots,G\}$ be their estimates, either from the proposed method or the rotation method. \zx{When data are generated from an approximate bi-factor model, the true group clusters  $\{\mathcal{B}_{g}^{*}, g=1, \ldots, G\}$ are based on the corresponding bi-factor loading matrix before the perturbation.} As the group factors can only be recovered up to label swapping, as Theorem~\ref{thm:id} suggests, we measure the matching between the true and estimated structure up to a permutation of the factor labels. Specifically, the following two evaluation criteria are considered: 
\begin{itemize}
    \item Exact Match Criterion (EMC): $\max_{\sigma \in \tilde{\mathcal {P}}}\prod_{g=1}^{G}\mathbf{1}(\mathcal{B}_{\sigma(g)} = \mathcal{B}_{g}^{*})$, which equals 1 when the bi-factor structure is correctly learned and 0 otherwise. Here, $(\sigma(1), ..., \sigma(G))$ is a permutation of $1, ..., G$, and $\tilde{\mathcal{P}}$ is the set of all such permutations. 
     
    \item Average Correctness Criterion (ACC): $\max_{\sigma\in\tilde{\mathcal{P}}}\sum_{g=1}^{G}(\vert \mathcal{B}_{\sigma(g)}\cap \mathcal{B}_{g}^{*}\vert + \vert\mathcal{B}_{\sigma(g)}^{C} \cap \mathcal{B}_{g}^{*C}\vert) / (JG)$, \jw{which is the 
    proportion of correctly identified zero and nonzero group loadings.}
    Here for any set $\mathcal{B}$, $\mathcal{B}^{C} = \{1,\ldots,J\} \setminus \mathcal{B}$ is the complement of set $\mathcal{B}$.
\end{itemize}

Here, the EMC measures the perfect recovery of the true bi-factor structure, while the ACC can be viewed as a smooth version of EMC
that measures the level of partial recovery. EMC = 1 when ACC = 1 and, EMC = 0 when ACC $< 1$. A larger value of ACC indicates a higher level of partial recovery of the true bi-factor structure. \jw{More specifically, for a given permutation $\sigma \in \tilde{\mathcal{P}}$, the quantity
$\vert \mathcal{B}_{\sigma(g)}\cap \mathcal{B}_{g}^{*}\vert + \vert\mathcal{B}_{\sigma(g)}^{C} \cap \mathcal{B}_{g}^{*C}\vert$ 
computes the number of correctly identified nonzero and zero loadings for group factor $g$. For example, consider a case with $J = 15$ items and $\mathcal{B}_{1}^{*} = \{1,4,7,10,13\}$ for the first group factor. If 
$\mathcal{B}_{\sigma(1)} = \mathcal{B}_{1}^{*}$, then for the first group factor, we have $\vert \mathcal{B}_{\sigma(1)}\cap \mathcal{B}_{1}^{*}\vert + \vert\mathcal{B}_{\sigma(1)}^{C} \cap \mathcal{B}_{1}^{*C}\vert = J = 15$, i.e., all the nonzero and zero loadings have been correctly identified. If, instead, $\mathcal{B}_{\sigma(1)} = \{1,2,7,10,13\}$, then $\vert \mathcal{B}_{\sigma(1)}\cap \mathcal{B}_{1}^{*}\vert + \vert\mathcal{B}_{\sigma(1)}^{C} \cap \mathcal{B}_{1}^{*C}\vert = 13$, i.e., 13 out of the 15 nonzero and zero loadings have been correctly identified. The quantity $\sum_{g=1}^{G}(\vert \mathcal{B}_{\sigma(g)}\cap \mathcal{B}_{g}^{*}\vert + \vert\mathcal{B}_{\sigma(g)}^{C} \cap \mathcal{B}_{g}^{*C}\vert) / (JG)$ thus computes the proportion of correctly identified zero and nonzero group loadings under the given permutation $\sigma$. The ACC considers all possible permutations of the group factor labels to account for the label indeterminacy. 
}

To examine the recovery of the bi-factor structure, \zx{we consider $(J,G) \in \{(15,3),(30,5)\}$ and $N \in \{500,2000\}$. These choices, combined with the two settings for the data generation mechanism, lead to eight simulation settings. For each setting, we let 
$\mathcal{B}_{g}^{*} = \{g,g+G ,\ldots,g + G(J/G-1)\}$ for $g = 1,\ldots, G$, $\Psi^* = \mathbb{I}_{J\times J}$, and $\Phi^* = \Phi^{*}(\boldsymbol \gamma^{*})$ follow the reparameterization in Section \ref{subsec:alm}, where the entries of $\boldsymbol \gamma^{*}$ are i.i.d., following a Uniform$(-0.5,0.5)$ distribution. Under the settings where data are generated from an exact bi-factor model, we generate the true loading matrix $\Lambda^*$ by 
\begin{equation}\label{eq:sim case1}
\begin{aligned}
\lambda_{jk}^{*} = \left\{
    \begin{array}{l}
        u_{jk} \mbox{~~if~~} k=1, \\
        0 \mbox{~~if~~} k>1 ,j \notin \mathcal{B}_{k-1}^{*}, \\
        {(1-2x_{jk})u_{jk}} \mbox{~~if~~} k>1, j\in\mathcal{B}_{k-1}^{*},
    \end{array}
\right.
\end{aligned}
\end{equation}
for $j = 1,\ldots,J$ and $k = 1,\ldots, G+1$. In \eqref{eq:sim case1}, $u_{jk}$s are i.i.d., following a Uniform$(0.2,1)$ distribution, and $x_{jk}$s are i.i.d., following a Bernoulli$(0.5)$ distribution. 
Under the settings where data are generated from an approximate bi-factor model, we generate the true loading matrix $\Lambda^*$ by 
\begin{equation}\label{eq:sim case2}
\begin{aligned}
\lambda_{jk}^{*} = \left\{
    \begin{array}{l}
        u_{jk} \mbox{~~if~~} k=1, \\
        (1-2x_{jk})w_{jk} \mbox{~~if~~} k>1 ,j \notin \mathcal{B}_{k-1}^{*}, \\
        (1-2x_{jk})u_{jk} \mbox{~~if~~} k>1, j\in\mathcal{B}_{k-1}^{*},
    \end{array}
\right.
\end{aligned}
\end{equation}
for $j = 1,\ldots,J$ and $k = 1,\ldots, G+1$. Here, $u_{jk}$s and $x_{jk}$s are generated in the same way as those in the exact bi-factor model, and $w_{jk}$ are i.i.d., following a Uniform$(0,0.1)$ distribution. In \eqref{eq:sim case2}, the nonzero values of $\lambda_{jk}^{*}$ when $k>1$ and $j \notin \mathcal{B}_{k-1}^{*}$ represent the perturbation of $\Lambda^*$ from an exact bi-factor structure. 
}

For each setting, we first generate $\Lambda^{*}$ and  $\Phi^{*}$ once and use them to generate 100 datasets. 
\zx{The true parameter values for these simulations are given in Appendix \ref{appen:population}. The results about the estimation of the loading matrix are shown in Table~\ref{tab:std1mse}. When data are generated from an exact bi-factor model, the proposed method outperforms the rotation method in terms of the MSE of the estimated loading matrix, as shown in Table~\ref{tab:std1mse}(a). When data are generated from an approximate bi-factor model, as shown in Table~\ref{tab:std1mse}(b), 
the proposed method is slightly better under the small-sample settings when $N=500$ but slightly worse under the large-sample settings when $N = 2000$. The disadvantage of the proposed method under the large-sample settings is due to the bias brought by model misspecification. That is, the data generation model is not an exact bi-factor model, while the proposed method restricts its estimates in the space of exact bi-factor models. }

\zx{The results about the recovery of the bi-factor structure based on the EMC and ACC metrics are shown in Tables~\ref{tab:std1emc} and~\ref{tab:std1acc}, respectively. For the rotation method, the threshold $\delta = 0.2$ yields the best results among the three threshold choices under all the simulation settings and for both performance metrics. However, even the results of the rotation method under this choice of threshold are not as good as those from the proposed method, especially when we look at the EMC metric. For example, when $J = 30$, $G=5$, and $N=500$, the rotation method with $\delta = 0.2$ can only correctly recover the entire bi-factor structure 15 times among 100 simulations, while the proposed method can correctly recover it 85 times. }

\begin{table}[htbp]
\centering
\caption{\zx{Simulation results of the MSE of $\hat{\Lambda}$ estimated by the proposed ALM method and the exploratory bi-factor rotation method.}}
\begin{subtable}[t]{0.45\textwidth}
\centering
    \begin{tabular}{ |c |c |c |c| }
    \hline
    $(J,G)$&$N$ & ALM & Rotation\\
    \hline
    (15,3) & 500 & $2.10\times 10^{-3}$ & $3.62\times 10^{-3}$  \\
    \hline
    &2000 & $0.54\times 10^{-3}$ & $0.92\times 10^{-3}$   \\
    \hline
    (30,5) & 500 & $1.36\times 10^{-3}$ & $4.94\times 10^{-3}$  \\
    \hline
    &2000 & $0.30\times 10^{-3}$ & $1.15\times 10^{-3}$  \\
    \hline

    \end{tabular}
    \vspace{0.1cm}
\caption{The exact bi-factor model cases.}
\label{tab:std1mse1}
\end{subtable}
\hfill
\begin{subtable}[t]{0.45\textwidth}
\centering
\begin{tabular}{ |c| c| c| c| }
    \hline
    $(J,G)$&$N$ & ALM & Rotation \\
    \hline
    (15,3) & 500 & $4.74\times 10^{-3}$ & $5.92\times 10^{-3}$  \\
    \hline
    &2000 & $3.06\times 10^{-3}$ & $2.40\times 10^{-3}$   \\
    \hline
    (30,5) & 500 & $3.75\times 10^{-3}$ & $3.88\times 10^{-3}$   \\
    \hline
    &2000 & $2.63\times 10^{-3}$ & $1.25\times 10^{-3}$  \\
    \hline

    \end{tabular}
    \vspace{0.1cm}
\caption{The approximate bi-factor model cases.}
\label{tab:std1mse2}
\end{subtable}
\label{tab:std1mse}
\end{table}

\begin{table}[htbp]
\centering
\caption{\zx{Simulation results of the EMC of the proposed ALM method and the exploratory bi-factor rotation method with three choices of hard thresholding parameter $\delta$.}}
\begin{subtable}[t]{0.6\textwidth}
\centering
    \begin{tabular}{ |c| c| c| c| c |c|  }
    \hline
    $(J,G)$&$N$ & ALM & $\delta = 0.1$&$\delta = 0.2$& $\delta = 0.3$ \\
    \hline
    (15,3) & 500 & 1.00 & 0.18 & 0.90 & 0.28 \\
    \hline
    &2000 & 1.00 & 0.99 & 1.00 & 0.50  \\
    \hline
    (30,5) & 500 & 0.85 & 0.00 & 0.15 &  0.00  \\
    \hline
    &2000 & 1.00 & 0.55 & 0.68 & 0.00 \\
    \hline

    \end{tabular}
\vspace{0.1cm}
\caption{The exact bi-factor model cases.}
\label{tab:std1emc1}
\vspace{0.1cm}
\end{subtable}
\hfill
\begin{subtable}[t]{0.6\textwidth}
\centering
\begin{tabular}{ |c| c| c| c| c| c|  }
    \hline
    $(J,G)$&$N$ & ALM & $\delta = 0.1$&$\delta = 0.2$& $\delta = 0.3$ \\
    \hline
    (15,3) & 500 & 0.99 & 0.00 & 0.62 & 0.33 \\
    \hline
    &2000 & 1.00 & 0.02 & 0.94 & 0.67 \\
    \hline
    (30,5) & 500 & 0.86 & 0.00 & 0.23 & 0.00 \\
    \hline
    &2000 & 1.00 & 0.00 & 0.82 & 0.00  \\
    \hline

    \end{tabular}
\vspace{0.1cm}
\caption{The approximate bi-factor model cases.}
\label{tab:std1emc2}
\vspace{0.1cm}
\end{subtable}
\label{tab:std1emc}
\end{table}

\begin{table}[htbp]
\centering
\caption{\zx{Simulation results of the ACC of the proposed ALM method and the exploratory bi-factor rotation method with three choices of hard thresholding parameter $\delta$.}}
\begin{subtable}[t]{0.6\textwidth}
\centering
    \begin{tabular}{ |c| c| c| c| c| c|  }
    \hline
    $(J,G)$&$N$ & ALM & $\delta = 0.1$&$\delta = 0.2$& $\delta = 0.3$ \\
    \hline
    (15,3) & 500 & 1.000 & 0.966 & 0.998 & 0.976 \\
    \hline
    &2000 & 1.000 & 0.999 & 1.000 & 0.987 \\
    \hline
    (30,5) & 500 & 0.998 & 0.892 & 0.987 & 0.973 \\
    \hline
    &2000 & 1.000 & 0.996 & 0.998 & 0.980  \\
    \hline

    \end{tabular}
\vspace{0.1cm}
\caption{The exact bi-factor model cases.}
\label{tab:std1acc1}
\vspace{0.1cm}
\end{subtable}
\hfill
\begin{subtable}[t]{0.6\textwidth}
\centering
\begin{tabular}{ |c| c| c| c| c| c|  }
    \hline
    $(J,G)$&$N$ & ALM & $\delta = 0.1$&$\delta = 0.2$& $\delta = 0.3$ \\
    \hline
    (15,3) & 500 & 0.999 & 0.864 & 0.989 & 0.978 \\
    \hline
    &2000 & 1.000 & 0.928 & 0.999 & 0.991 \\
    \hline
    (30,5) & 500 & 0.998 & 0.848 & 0.989 & 0.978 \\
    \hline
    &2000 & 1.000 & 0.927 & 0.999 & 0.980  \\
    \hline

    \end{tabular}
\vspace{0.1cm}
\caption{The approximate bi-factor model cases.}
\label{tab:std1acc2}
\vspace{0.1cm}
\end{subtable}
\label{tab:std1acc}
\end{table}

\subsection{Study II}\label{subsec:study 2}
In this study, we examine the selection of the number of factors by $\text{BIC}_{G}$ in Section~\ref{subsec:selection}. We compare it with 
selecting the number of factors under the exploratory factor analysis model without assuming a bi-factor structure. For the proposed method, we set the candidate set $\mathcal G = \{G^{*}-1, G^{*}, G^{*}+1\}$, where $G^*$ is the true number of group factors. 
For exploratory factor analysis, we also use the BIC for determining the number of factors, which is defined as 
$$
\text{BIC}_{K}^{e} = l_{K}^{e} + (JK - K(K-1)/2)\log(N),
$$
where $K$ is the number of factors in the exploratory factor analysis model, and 
$l_{K}^{e} = l(\hat \Lambda \hat \Lambda^{\top}+\Psi(\hat \psi);S)$ with $\hat \Lambda$ and $\hat \psi$ from \eqref{eq:efa}. As the number of factors in the exploratory factor analysis model equals the number of group factors plus one, we choose $K$ from the candidate set $\mathcal K = \{G+1: G\in \mathcal G\}$. Let $\hat K =  \argmin_{K\in \mathcal K} ~\text{BIC}_K^e.$ Then the estimate of $G$ by exploratory factor analysis is $\hat G = \hat K-1$. The selection accuracy is evaluated by the 
selection correctness (SC) criterion, defined as  $\mathbf{1}(\hat{G} = G^*)$, where $\hat G$ is obtained using the proposed method in Section~\ref{subsec:selection} or under the exploratory factor analysis model described above.

We conduct simulations under four settings, with $(J,G^*) \in \{(15,3),(30,5)\}$ and $N \in \{500,2000\}$ and the data generation models being the same exact bi-factor models in Study I. For each setting, 100 independent simulations are performed. The results are given in Table~\ref{tab:std2}, where the column indicated by $\bar G$ reports the average value of $\hat G$. 
We see that both methods can select the number of factors reasonably well, with their accuracy being 100\% when $G^* = 3$ for both sample sizes. {When $G^* = 5$ and the sample size $N=2000$, the proposed method achieves an accuracy of 100\%, and the exploratory factor analysis method achieves an accuracy of 99\%}. This is not surprising, given that the BIC has asymptotic consistency in 
selecting the number of factors under both models. It is worth noting that, when $G^* = 5$ and for the smaller sample size $N=500$, which is the most challenging setting, the proposed method achieves an accuracy of 98\%, {while that of the exploratory factor analysis method is zero. More precisely, the exploratory factor analysis method selects $G =4$ in all the replications.} It suggests that the proposed method has an advantage in smaller sample settings. This result 
is expected, as the exploratory factor analysis method doesn't utilize the information about the bi-factor structure. Consequently, it 
overestimates the number of parameters, which leads to a larger penalty term and, subsequently, a tendency to under-select $G$.
\begin{table}[h]
    \centering
    \caption{Simulation results of the selection of the number of factors by BIC.}
    \label{tab:std2}
    \begin{tabular}{ c  c  c  c c c }
    \hline
     & & \multicolumn{2}{c}{ALM} & \multicolumn{2}{c}{Exploratory}\\
    \hline
    $(J,G^*)$ & $N$ & $\bar{G}$ & SC & $\bar{G}$ & SC\\
    \hline
    (15,3) & 500 & 3 & 1 & 3 & 1\\
    \hline
     &2000 &3& 1 & 3 & 1 \\
    \hline
    (30,5) & 500 & 5.02 & 0.98 & 4 & 0\\
    \hline
     &2000 & 5 & 1 & 4.99 & 0.99 \\
    \hline

    \end{tabular}
\end{table}

\section{Real Data Analysis}\label{sec:real data}
In this section, we apply the exact exploratory bi-factor analysis to a personality assessment dataset based on the
International Personality Item Pool (IPIP) NEO 120 personality inventory \citep{johnson2014measuring}\footnote{The data are downloaded from https://osf.io/tbmh5/}. We investigate the structure of the Extraversion scale based on a sample of 1,107 UK male participants aged between 25 and 30 years. This scale consists of 24 items, which are designed to measure six facets of Extraversion, including Friendliness (E1), Gregariousness (E2), Assertiveness (E3), Activity Level(E4), Excitement-Seeking (E5) and Cheerfulness (E6); see Section~\ref{appen:ItemKey} for the details. 
All the items are on a 1-5 Likert scale, and the reversely worded items have been reversely scored so that a larger score always means a higher level of extraversion. There is no missing data. 
Detailed descriptions of the items can be found in
the Appendix~\ref{appen:ItemKey}.

Using a candidate set $\mathcal{G} = \{2,\ldots,12\}$, the BIC procedure given 
in Section \ref{subsec:selection} selects seven group factors. The estimated loading matrix is given in Table~\ref{tab:loading}, and the estimated factor correlation matrix is given below. \jw{The estimated model fits the data well, as implied by the commonly used fit statistics, including RMSEA = 0.044, SRMR = 0.031, CFI = 0.965, and TLI = 0.953.} We point out that the estimated bi-factor structure does not meet Condition~\ref{cond:altersep}, one of the sufficient conditions for Theorem~\ref{thm:id}. However, as shown in Appendix~\ref{appen:Proofrd}, with some additional mild assumptions, this structure and its parameters are still identifiable. 

We now examine the estimated model. 
We first notice that the loadings on the general factor are all positive. 
Consequently, this factor can be easily interpreted as the general extraversion factor. The seven group factors are closely related to the six aspects of extraversion. Specifically, we interpret the group factors G1, G3, G4 and G5 as the Friendliness, Cheerfulnes, Assertiveness, and Activity Level factors, respectively, as the items loading on them highly overlap with the items that are used to define the corresponding aspects. In particular, the items loading on G3 and G5 are exactly those that define the Cheerfulness and Activity Level aspects, respectively. The items loading on G1 include all the items that define the Friendliness aspect and an additional item ``7. Prefer to be alone", a negatively worded item that is used to define the Gregariousness aspect. This additional item aligns well with the Friendliness dimension, given the social nature behind it. In addition, the items loading on G4 consist of all the items that define the Assertiveness aspect and an additional item ``6. Talk to a lot of different people at parties", which is used to define the Gregariousness aspect. This additional item aligns with the Assertiveness dimension in that talking to many different people at parties typically requires sufficient confidence, 
a key element of Assertiveness. 

\begin{table}[h]
    \centering
    \caption{Estimated bi-factor loading matrix $\hat{\Lambda}$ with seven group factors. }
    \label{tab:loading}
    \begin{tabular}{| r | r |r | r | r | r | r | r | r | r |}
    \hline
    Items & Sign & General & G1 & G2 & G3 & G4
    & G5 & G6 & G7\\
    \hline
        1& +E1&0.85 & 0.26 & 0 & 0 & 0 & 0 & 0 & 0 \\ 
        2& +E1 &0.73 & 0.48 & 0 & 0 & 0 & 0 & 0 & 0 \\ 
        3& $-$E1&0.74 & 0.57 & 0 & 0 & 0 & 0 & 0 & 0 \\ 
        4& $-$E1&0.68 & 0.58 & 0 & 0 & 0 & 0 & 0 & 0 \\         
        5& +E2&0.94 & 0 & 0 & 0 & 0 & 0 & 0.26 & 0 \\ 
        6& +E2&1.01 & 0 & 0 & 0 & 0.17 & 0 & 0 & 0 \\ 
        7& $-$E2&0.53 & 0.52 & 0 & 0 & 0 & 0 & 0 & 0 \\ 
        8& $-$E2&0.67 & 0 & 0 & 0 & 0 & 0 & 1.06 & 0 \\ 
        9& +E3&0.37 & 0 & 0 & 0 & 0.86 & 0 & 0 & 0 \\ 
        10& +E3&0.38 & 0 & 0 & 0 & 0.81 & 0 & 0 & 0 \\ 
        11& +E3&0.28 & 0 & 0 & 0 & 0.74 & 0 & 0 & 0 \\
        12& $-$E3&0.39 & 0 & 0 & 0 & 0.75 & 0 & 0 & 0 \\ 
        13& +E4&0.20 & 0 & 0 & 0 & 0 & 0.81 & 0 & 0 \\ 
        14& +E4&0.40 & 0 & 0 & 0 & 0 & 0.82 & 0 & 0 \\ 
        15& +E4&0.40 & 0 & 0 & 0 & 0 & 0.60 & 0 & 0 \\ 
        16& $-$E4&0.04 & 0 & 0 & 0 & 0 & 0.47 & 0 & 0 \\
        17& +E5&0.46 & 0 & 0 & 0 & 0 & 0 & 0 & 0.46 \\ 
        18& +E5&0.47 & 0 & 0 & 0 & 0 & 0 & 0 & 0.71 \\ 
        19& +E5&0.35 & 0 & 0.86 & 0 & 0 & 0 & 0 & 0 \\ 
        20& +E5&0.56 & 0 & 0.71 & 0 & 0 & 0 & 0 & 0 \\ 
        21& +E6&0.59 & 0 & 0 & 0.42 & 0 & 0 & 0 & 0 \\ 
        22& +E6&0.64 & 0 & 0 & 0.48 & 0 & 0 & 0 & 0 \\ 
        23& +E6&0.46 & 0 & 0 & 0.76 & 0 & 0 & 0 & 0 \\ 
        24& +E6&0.41 & 0 & 0 & 0.74 & 0 & 0 & 0 & 0 \\ \hline
    \end{tabular}
\end{table}

\begin{equation}
\begin{aligned}
\hat{\Phi} = \left(
\begin{matrix}
1 & 0& 0 & 0& 0 & 0 & 0 & 0 \\
0 & 1 & -0.24 & 0.54 & 0.37 & 0.16 & 0.51 & 0.08 \\ 
0 & -0.24 & 1 & -0.04 & 0.05 & -0.09 & -0.01 & 0.51 \\ 
0 & 0.54 & -0.04 & 1 & 0.30 & 0.28 & 0.20 & 0.25 \\ 
0 & 0.37 & 0.05 & 0.30 & 1 & 0.38 & 0.15 & 0.29 \\ 
0 & 0.16 & -0.09 & 0.28 & 0.38 & 1 & 0.11 & 0.22 \\ 
0 & 0.51 & -0.01 & 0.20 & 0.15 & 0.11 & 1 & 0.10 \\ 
0 & 0.08 & 0.51 & 0.25 & 0.29 & 0.22 & 0.10 & 1 \\ 
\end{matrix}
\right).
\end{aligned}
\end{equation}

The group factors G2 and G7 may be viewed as two subdimensions of the Excitement-Seeking aspect, as each of them is loaded with two items that define the Excitement-Seeking aspect. Specifically, G2 is loaded with the items ``19. Enjoy being reckless" and ``20. Act wild and crazy", while G7 is loaded with the items ``17. Love excitement" and ``18. Seek adventure". We believe that G2 
emphasises the thrill of the moment of excitement and the disregard for potential consequences, while G7 emphasizes the pursuit of meaningful and fulfilling experiences. Therefore, we interpret G2 as the Reckless Excitement-Seeking factor, while interpret G7 as the Meaningful Excitement-Seeking factor. Finally, G6 is loaded with two items, ``5. Love large parties" and ``8. Avoid crowds", where item 8 is reversely worded. Both items are used to define the Gregariousness aspect. Compared with items 6 and 7, which are also used to define the Gregariousness aspect but now load on two different group factors, these two items may better reflect the essence of Gregariousness -- the tendency to enjoy the company of others. We thus interpret G6 as the Gregariousness factor. We also notice that most correlations between the group factors are positive, except for some of the correlations with G2. Specifically, we see that G2 (Reckless Excitement-Seeking) has a moderate negative correlation with G1 (Friendliness) while a reasonably high correlation with G7 (Meaningful Excitement-Seeking). 

\yc{We have also applied the bi-factor rotation method of \cite{jennrich2012exploratory} to the same data, which gives a solution with seven group factors. The resulting bi-factor structure is similar to that given by the proposed method, except that 
the rotation solution does not seem to contain a clear Friendliness factor; see Appendix~\ref{appen:real data rotation} for further details.}

\section{Discussions} \label{sec:diss}

This paper proposes a constraint-based optimisation method for exploratory bi-factor analysis. This method turns the problem of exploratory bi-factor analysis into an equality-constrained optimisation problem in a continuous domain and solves this optimisation problem by an augmented Lagrangian method. Compared with the rotation method of \cite{jennrich2011exploratory,jennrich2012exploratory}, the proposed method can learn an exact loading structure without a post-hoc treatment step. In the simulation studies, the ALM method achieves higher estimation accuracy when data are generated from an exact bi-factor model. In addition, it has a higher chance of recovering the true bi-factor structure than the rotation method, whether data are generated from an exact or approximate bi-factor model. Moreover, the ALM method correctly estimates the number of the group factors in most of the simulation replications. 
In the real data analysis concerning an Extraversion personality scale, the ALM method yields a bi-factor structure with seven group factors. The identified group factors are psychologically interpretable. 

An innovation of current research is turning a model selection problem, which is combinatory by nature, into a continuous optimization problem. This avoids a computationally intensive search procedure for fitting many possible models and comparing their fits, noting that the number of possible models grows exponentially with $J$. We admit that this continuous optimization formulation also has a limitation. The space for the bi-factor loading matrix characterised by the nonlinear equality constraints in \eqref{eq:constraint} is highly nonconvex, and thus, the ALM may sometimes converge to a local minimum. To alleviate this issue, we suggest running the ALM with multiple random starting points and then choosing the solution with the smallest objective function value. Based on our simulation results, using 50 starting points seems sufficient to converge to somewhere close to the true parameters up to a label swapping of the group factors and a sign indeterminacy of loadings in almost all replications under the settings considered in the simulation study.

This research leads to several new directions for exploratory analysis of factor models with structure constraints on the loading matrix. First, as pointed out earlier, the proposed approach can be easily adapted to non-linear bi-factor models for dichotomous, ordinal, and nominal data. Under the 
confirmatory setting, these models are typically estimated by maximising the marginal log-likelihood function or other objective functions (e.g., a composite likelihood). Under the exploratory setting, one only needs to maximise the same objective function subject to the same bi-factor constraints in 
 \eqref{eq:constraint}, for which the ALM adapts naturally. 
It is worth noting that, however, the marginal likelihood of the non-linear bi-factor models typically involves multidimensional integrals with respect to the factors, and they do not have an analytic form. Consequently, solving the Lagrangian augmented objective functions using the standard expectation-maximisation (EM) algorithm \citep{dempster1977maximum,bock1981marginal} can be computationally intensive. One possible solution is to use a stochastic approximation method \citep{zhang2022computation,oka2024learning}. These methods avoid the high computational cost of numerical integrals in the expectation-maximisation algorithm by constructing stochastic gradients of the marginal log-likelihood through Markov chain Monte Carlo sampling. 

Second, the proposed constraints can also be combined with exploratory factor analysis techniques to learn a bi-factor structure in two steps. Suppose an initial loading matrix estimate $\hat \Lambda$ has been obtained under the constraint that the factors are orthogonal (i.e., $\Phi$ is an identity matrix). 
It may be obtained by a standard exploratory factor analysis method. 
In that case, we can find a bi-factor structure that best approximates $\hat \Lambda$ (up to a rotation) by minimising 
$\Vert\Lambda \Phi(\boldsymbol \gamma) \Lambda^\top - \hat \Lambda (\hat \Lambda)^\top\Vert_F$ with respect to $\Lambda$ and $\boldsymbol{\gamma}$
under the constraints in \eqref{eq:constraint}.
This optimisation can again be solved by an ALM. 

Third, as we demonstrate in Appendix~\ref{appen:hfm}, the set of constraints in \eqref{eq:constraint} can be extended to characterise the loading structure of a hierarchical factor model \citep{schmid1957development,yung1999relationship}, which can be used to learn a hierarchical factor structure. This exploratory hierarchical factor analysis may allow researchers to learn more refined and interpretable latent structures from psychometric data. However, one should note that exploratory hierarchical factor analysis is more complex than exploratory bi-factor analysis, as the factor hierarchy in the former can be much more complex than the two-layer hierarchy in the latter. 
The learning algorithm in Appendix~\ref{appen:hfm} requires the factor hierarchy to be known (see, e.g., Panel (b) of Figure~\ref{fig:hier}). The problem becomes more challenging when the factor hierarchy is unknown, in which case we need to learn both the factor hierarchy and the loading pattern of the variables. We leave this problem for future investigation.} 

Finally, we point out that the proposed method always returns an estimated bi-factor model, whether it fits the data or not.
The simulation study in Section~\ref{sec:sim} shows that the proposed method has robust performance when data are generated by an approximate bi-factor model. However, under more general settings, it remains to test the goodness-of-fit of the estimated model to decide whether a bi-factor model suits the data. If the bi-factor model does not fit the data well, we may consider a more flexible factor model. For example, we may apply the bi-factor rotation method or 
a rotation method for traditional exploratory factor analysis to allow for more cross-loadings. Alternatively, we may learn approximate bi-factor models in an exploratory manner by relaxing the equality constraints in \eqref{eq:optimisation} with inequality constraints in the form of $|\lambda_{jk}\lambda_{jk'}|\leq \epsilon$, for all $j = 1, ..., J$, and $k, k' = 2, ..., G+1$, $k\neq k'$, where $\epsilon$ is a tuning parameter that controls the level of approximation to a bi-factor model.  A larger value of $\epsilon$ leads to a more flexible model space and, thus, 
a more satisfactory fit,
while a smaller value of $\epsilon$ leads to a better approximation to a bi-factor model that may have better interpretability. In this sense,  $\epsilon$ provides a trade-off between model goodness-of-fit and bi-factor interoperability. This inequality-constrained optimisation may be solved using an interior-point method, which can incorporate the inequality constraints through suitable barrier functions (e.g., log-barriers). We leave this idea for future investigation.

\vspace{2cm}
\appendix
\noindent
{\Large Appendix}
\numberwithin{equation}{section}
\numberwithin{figure}{section}
\numberwithin{table}{section}

\section{Reparameterization of $\Phi$}\label{appen:repara}
To deal with the constraints in $\Phi$, we consider the following reparameterisation that has been considered in \cite{alfonzetti2024pairwise}, which is also similar to the implementation in the state-of-the-art statistical software Stan \citep{stan}: 

\begin{equation}
\begin{aligned}
\Phi = \left( \begin{bmatrix}
    1 & \mathbf{0}^\top \\
    \mathbf{0} & U^{T} 
\end{bmatrix} \right)
\left( \begin{bmatrix}
    1 & \mathbf{0}^\top \\
    \mathbf{0} & U 
\end{bmatrix} \right),
\end{aligned}
\end{equation}
where $U$ is defined recursively by 
\begin{equation}
\begin{aligned}
U_{ij} = \left\{
    \begin{array}{l}
        0 \mbox{~~if~~} i>j; \\
        1 \mbox{~~if~~} i=j=1;\\
        z_{ij} \mbox{~~if~~} 1=i<j;\\
        \frac{z_{ij}}{z_{(i-1)j}}U_{(i-1)j}(1-z_{(i-1)j}^{2})^{1/2} \mbox{~~if~~} 1<i<j;\\
        \frac{U_{(i-1)j}}{z_{(i-1)j}}(1-z_{(i-1)j}^{2})^{1/2} \mbox{~~if~~} 1<i=j.
    \end{array}
\right.
\end{aligned}
\end{equation}
Here $z_{ij} = \tanh(\gamma_{ij})$ is the Fisher's transformation of $G(G-1)/2$ unconstrained parameters $\gamma_{ij}$. 

\section{Population Parameter Values in Simulations}\label{appen:population}
In this section, we supplement the population values of factor loadings and factor correlations in Section~\ref{sec:sim}. Under the setting $(J,G) = (15,3)$, the loading matrix $\Lambda^*$ and $\Phi^*$ in \eqref{eq:sim case1} and \eqref{eq:sim case2} are given in \eqref{eq:G3 loading 1}, \eqref{eq:G3 Psi 1}, \eqref{eq:G3 loading 2} and \eqref{eq:G3 Psi 2} respectively.
\begin{align}\label{eq:G3 loading 1}
    \Lambda^* &= \left(\begin{matrix}
        0.36 & -0.60 & 0 & 0 \\ 
        0.89 & 0 & 0.64 & 0 \\ 
        0.94 & 0 & 0 & 0.92 \\ 
        0.90 & 0.33 & 0 & 0 \\ 
        0.45 & 0 & -0.90 & 0 \\ 
        0.24 & 0 & 0 & 0.86 \\ 
        0.45 & -0.60 & 0 & 0 \\ 
        0.74 & 0 & -0.75 & 0 \\ 
        0.64 & 0 & 0 & -0.60 \\ 
        0.50 & -0.83 & 0 & 0 \\ 
        0.54 & 0 & -0.43 & 0 \\ 
        0.43 & 0 & 0 & 0.86 \\ 
        0.46 & -0.33 & 0 & 0 \\ 
        0.67 & 0 & 0.34 & 0 \\ 
        0.90 & 0 & 0 & -0.83 \\ 
    \end{matrix}\right)    
\end{align}

\begin{align}\label{eq:G3 Psi 1}
    \Phi^* &= \left(\begin{matrix}
        1 & 0 & 0 & 0 \\ 
        0 & 1 & 0.29 & 0.11 \\ 
        0 & 0.29 & 1 & 0.01 \\ 
        0 & 0.11 & 0.01 & 1 \\ 
        \end{matrix}\right)
\end{align}

\begin{align}\label{eq:G3 loading 2}
    \Lambda^* &= \left(\begin{matrix}
        0.36 & -0.6 & -0.05 & 0.06 \\ 
        0.89 & 0.04 & 0.64 & 0.09 \\ 
        0.94 & -0.09 & -0.08 & 0.92 \\ 
        0.90 & 0.33 & 0.02 & -0.09 \\ 
        0.45 & -0.07 & -0.90 & -0.06 \\ 
        0.24 & -0.02 & 0.09 & 0.86 \\ 
        0.45 & -0.60 & -0.09 & 0.02 \\ 
        0.74 & 0.05 & -0.75 & 0.09 \\ 
        0.64 & 0.05 & -0.02 & -0.60 \\ 
        0.50 & -0.83 & 0.05 & 0.06 \\ 
        0.54 & 0.02 & -0.43 & 0.07 \\ 
        0.43 & 0.03 & -0.09 & 0.86 \\ 
        0.46 & -0.33 & 0.00 & 0.04 \\ 
        0.67 & 0.05 & 0.34 & 0.04 \\ 
        0.90 & 0.00 & -0.06 & -0.83 \\ 
    \end{matrix}\right)    
\end{align}

\begin{align}\label{eq:G3 Psi 2}
    \Phi^* &= \left(\begin{matrix}
        1 & 0 & 0 & 0 \\ 
        0 & 1 & -0.22 & -0.07 \\ 
        0 & -0.22 & 1 & 0.46 \\ 
        0 & -0.07 & 0.46 & 1 \\ 
    \end{matrix}\right)    
\end{align}

Under the setting $(J,G) = (30,5)$, the loading matrix $\Lambda$ in \eqref{eq:sim case1} and \eqref{eq:sim case2} are given in \eqref{eq:G5 loading 1} and \eqref{eq:G5 loading 2}. The correlation matrix $\Phi$ in \eqref{eq:sim case1} and \eqref{eq:sim case2} are given in \eqref{eq:G5 Psi 1} and \eqref{eq:G5 Psi 2}.
\begin{equation}
\label{eq:G5 loading 1}
\footnotesize
    \Lambda^* = \left(\begin{array}{cccccc}
        0.40 & -0.90 & 0 & 0 & 0 & 0 \\ 
        0.53 & 0 & -0.62 & 0 & 0 & 0 \\ 
        0.21 & 0 & 0 & -0.78 & 0 & 0 \\ 
        0.69 & 0 & 0 & 0 & -0.81 & 0 \\ 
        0.57 & 0 & 0 & 0 & 0 & 0.90 \\ 
        0.91 & -0.93 & 0 & 0 & 0 & 0 \\ 
        0.31 & 0 & -0.47 & 0 & 0 & 0 \\ 
        0.41 & 0 & 0 & -0.79 & 0 & 0 \\ 
        0.31 & 0 & 0 & 0 & -0.35 & 0 \\ 
        0.90 & 0 & 0 & 0 & 0 & -0.59 \\ 
        0.59 & 0.80 & 0 & 0 & 0 & 0 \\ 
        0.83 & 0 & -0.96 & 0 & 0 & 0 \\ 
        0.84 & 0 & 0 & -0.89 & 0 & 0 \\ 
        0.29 & 0 & 0 & 0 & -0.23 & 0 \\ 
        0.90 & 0 & 0 & 0 & 0 & 0.77 \\ 
        0.68 & -0.56 & 0 & 0 & 0 & 0 \\ 
        0.95 & 0 & 0.89 & 0 & 0 & 0 \\ 
        0.87 & 0 & 0 & 0.61 & 0 & 0 \\ 
        0.45 & 0 & 0 & 0 & 0.41 & 0 \\ 
        0.52 & 0 & 0 & 0 & 0 & 0.51 \\ 
        0.43 & -0.42 & 0 & 0 & 0 & 0 \\ 
        0.43 & 0 & -0.30 & 0 & 0 & 0 \\ 
        0.34 & 0 & 0 & -0.31 & 0 & 0 \\ 
        0.73 & 0 & 0 & 0 & 0.53 & 0 \\ 
        0.69 & 0 & 0 & 0 & 0 & -0.83 \\ 
        0.30 & 0.34 & 0 & 0 & 0 & 0 \\ 
        0.25 & 0 & -0.60 & 0 & 0 & 0 \\ 
        0.58 & 0 & 0 & 0.37 & 0 & 0 \\ 
        0.28 & 0 & 0 & 0 & -0.28 & 0 \\ 
        0.85 & 0 & 0 & 0 & 0 & 0.82 \\ 
    \end{array}\right)    
\end{equation}

\begin{equation}
\label{eq:G5 loading 2}
\small
    \Lambda^* = \left(\begin{array}{cccccc}
        0.40 & -0.90 & -0.07 & 0.01 & 0.01 & 0.02 \\ 
        0.53 & 0.02 & -0.62 & -0.06 & 0.09 & 0.00 \\ 
        0.21 & 0.01 & -0.01 & -0.78 & -0.02 & -0.03 \\ 
        0.69 & -0.05 & 0.01 & 0.08 & -0.81 & -0.07 \\ 
        0.57 & 0.05 & 0.06 & -0.09 & 0.03 & 0.90 \\ 
        0.91 & -0.93 & -0.09 & -0.01 & 0.05 & 0.04 \\ 
        0.31 & 0.02 & -0.47 & 0.04 & 0.05 & -0.05 \\ 
        0.41 & 0.04 & -0.07 & -0.79 & 0.02 & 0.05 \\ 
        0.31 & 0.05 & 0.01 & -0.07 & -0.35 & -0.04 \\ 
        0.90 & 0.05 & -0.07 & -0.10 & 0.03 & -0.59 \\ 
        0.59 & 0.80 & -0.05 & -0.10 & -0.05 & 0.02 \\ 
        0.83 & 0.10 & -0.96 & -0.07 & 0.06 & -0.03 \\ 
        0.84 & 0.03 & 0.08 & -0.89 & -0.01 & -0.04 \\ 
        0.29 & 0.09 & -0.05 & -0.01 & -0.23 & 0.04 \\ 
        0.90 & -0.10 & -0.04 & 0.07 & -0.08 & 0.77 \\ 
        0.68 & -0.56 & 0.02 & -0.01 & -0.07 & -0.08 \\
        0.95 & 0.05 & 0.89 & -0.03 & 0.04 & -0.08 \\ 
        0.87 & 0.09 & 0.01 & 0.61 & -0.07 & 0.05 \\ 
        0.45 & 0.04 & -0.04 & 0.00 & 0.41 & 0.06 \\ 
        0.52 & -0.04 & -0.05 & -0.10 & -0.03 & 0.51 \\ 
        0.43 & -0.42 & -0.07 & -0.04 & -0.07 & -0.08 \\ 
        0.43 & 0.02 & -0.30 & 0.05 & -0.09 & 0.06 \\ 
        0.34 & 0.05 & -0.02 & -0.31 & 0.05 & -0.04 \\ 
        0.73 & -0.07 & 0.10 & 0.09 & 0.53 & 0.04 \\ 
        0.69 & -0.09 & -0.03 & -0.08 & -0.05 & -0.83 \\ 
        0.30 & 0.34 & 0.09 & -0.04 & 0.01 & 0.01 \\ 
        0.25 & 0.02 & -0.60 & 0.09 & 0.04 & 0.06 \\ 
        0.58 & -0.08 & -0.03 & 0.37 & -0.02 & 0.08 \\ 
        0.28 & -0.08 & 0.04 & -0.03 & -0.28 & -0.01 \\ 
        0.85 & -0.06 & 0.07 & 0.08 & -0.09 & 0.82 \\ 
     \end{array}\right)    
\end{equation}

\begin{align}\label{eq:G5 Psi 1}
    \Phi^* &= \left(\begin{matrix}
        1 & 0 & 0 & 0 & 0 & 0 \\ 
        0 & 1 & -0.37 & -0.26 & -0.41 & -0.07 \\ 
        0 & -0.37 & 1 & -0.08 & -0.14 & -0.26 \\
        0 & -0.26 & -0.08 & 1 & 0.42 & 0.06 \\ 
        0 & -0.41 & -0.14 & 0.42 & 1 & 0.20 \\ 
        0 & -0.07 & -0.26 & 0.06 & 0.20 & 1 \\ 
   \end{matrix}\right)    
\end{align}

\begin{align}\label{eq:G5 Psi 2}
    \Phi^* &= \left(\begin{matrix}
        1 & 0 & 0 & 0 & 0 & 0 \\ 
        0 & 1 & -0.15 & 0.15 & 0.10 & -0.19 \\ 
        0 & -0.15 & 1 & 0.22 & -0.01 & -0.03 \\ 
        0 & 0.15 & 0.22 & 1 & -0.03 & -0.15 \\ 
        0 & 0.10 & -0.01 & -0.03 & 1 & 0.28 \\ 
        0 & -0.19 & -0.03 & -0.15 & 0.28 & 1 \\ 
    \end{matrix}\right)    
\end{align}

\section{Sensitivity Analysis}\label{appen:sensitivity}
In this section, we carry out a sensitivity analysis on the parameters $c_{\theta}$ and $c_{\sigma}$ of the proposed ALM method. {We consider the same exact bi-factor model settings as in Study I of Section~\ref{subsec:study1}.} For each settings, we choose $c_{\theta}\in \{0.25,0.5,0.75\}$ and $c_{\sigma}\in\{5,10,15\}$, resulting in 9 possible combinations of $(c_{\theta},c_{\sigma})$. The estimation of loading matrix, the computation time of ALM, and the results of the recovery of the bi-factor structure are shown in Table~\ref{tab:SenMSE} to Table~\ref{tab:SenTime}. From the sensitivity analysis, we can see that the ALM's results are relatively stable with respect to the choice of parameters $c_{\theta}$ and $c_{\sigma}$.

\begin{table}[htbp]
\centering
\caption{Sensitivity Analysis of MSE of $\hat{\Lambda}$.}
\begin{subtable}[t]{0.6\textwidth}
\centering
\label{tab:Sen MSE1}
\begin{tabular}{|c|c c c|}
\hline
\diagbox{$c_{\sigma}$}{$c_{\theta}$} & 0.25 & 0.5 & 0.75 \\ 
\hline
5 & $2.10\times10^{-3}$ & $2.10\times10^{-3}$ & $2.10\times10^{-3}$ \\
10 & $2.10\times10^{-3}$ & $2.10\times10^{-3}$ & $2.10\times10^{-3}$ \\
15 & $2.10\times10^{-3}$ & $2.10\times10^{-3}$ & $2.10\times10^{-3}$ \\
\hline
\end{tabular}
\caption{$J=15$, $G=3$, $n=500$}
\vspace{0.2cm}
\end{subtable}
\begin{subtable}[t]{0.6\textwidth}
\centering
\label{tab:Sen MSE2}
\begin{tabular}{|c|c c c|}
\hline
\diagbox{$c_{\sigma}$}{$c_{\theta}$} & 0.25 & 0.5 & 0.75 \\ 
\hline
5 & $0.54\times10^{-3}$ & $0.54\times10^{-3}$ & $0.54\times10^{-3}$ \\
10 & $0.54\times10^{-3}$ & $0.54\times10^{-3}$ & $0.54\times10^{-3}$ \\
15 & $0.54\times10^{-3}$ & $0.54\times10^{-3}$ & $0.54\times10^{-3}$ \\
\hline
\end{tabular}
\caption{$J=15$, $G=3$, $n=2000$}
\vspace{0.2cm}
\end{subtable}
\begin{subtable}[t]{0.6\textwidth}
\centering
\label{tab:Sen MSE3}
\begin{tabular}{|c|c c c|}
\hline
\diagbox{$c_{\sigma}$}{$c_{\theta}$} & 0.25 & 0.5 & 0.75 \\ 
\hline
5 & $1.36\times10^{-3}$ & $1.38\times10^{-3}$ & $1.39\times10^{-3}$ \\
10 & $1.36\times10^{-3}$ & $1.42\times10^{-3}$ & $1.33\times10^{-3}$ \\
15 & $1.42\times10^{-3}$ & $1.36\times10^{-3}$ & $1.44\times10^{-3}$ \\
\hline
\end{tabular}
\caption{$J=30$, $G=5$, $n=500$}
\vspace{0.2cm}
\end{subtable}
\begin{subtable}[t]{0.6\textwidth}
\centering
\label{tab:Sen MSE4}
\begin{tabular}{|c|c c c|}
\hline
\diagbox{$c_{\sigma}$}{$c_{\theta}$} & 0.25 & 0.5 & 0.75 \\ 
\hline
5 & $0.30\times10^{-3}$ & $0.30\times10^{-3}$ & $0.30\times10^{-3}$ \\
10 & $0.30\times10^{-3}$ & $0.30\times10^{-3}$ & $0.30\times10^{-3}$ \\
15 & $0.30\times10^{-3}$ & $0.30\times10^{-3}$ & $0.30\times10^{-3}$ \\
\hline
\end{tabular}
\caption{$J=30$, $G=5$, $n=2000$}
\end{subtable}
\label{tab:SenMSE}
\end{table}

\begin{table}[htbp]
\centering
\caption{Sensitivity Analysis of EMC.}
\begin{subtable}[t]{0.45\textwidth}
\centering
\label{tab:Sen EMC1}
\begin{tabular}{|c|c c c|}
\hline
\diagbox{$c_{\sigma}$}{$c_{\theta}$} & 0.25 & 0.5 & 0.75 \\ 
\hline
5 & 1.00 & 1.00 & 1.00 \\
10 & 1.00 & 1.00 & 1.00 \\
15 & 1.00 & 1.00 & 1.00 \\
\hline
\end{tabular}
\caption{$J=15$, $G=3$, $n=500$}
\end{subtable}
\hfill
\begin{subtable}[t]{0.45\textwidth}
\centering
\label{tab:Sen EMC2}
\begin{tabular}{|c|c c c|}
\hline
\diagbox{$c_{\sigma}$}{$c_{\theta}$} & 0.25 & 0.5 & 0.75 \\ 
\hline
5 & 1.00 & 1.00 & 1.00 \\
10 & 1.00 & 1.00 & 1.00 \\
15 & 1.00 & 1.00 & 1.00 \\
\hline
\end{tabular}
\caption{$J=15$, $G=3$, $n=2000$}
\end{subtable}
\hfill
\begin{subtable}[t]{0.45\textwidth}
\centering
\label{tab:Sen EMC3}
\begin{tabular}{|c|c c c|}
\hline
\diagbox{$c_{\sigma}$}{$c_{\theta}$} & 0.25 & 0.5 & 0.75 \\ 
\hline
5 & 0.86 & 0.86 & 0.85 \\
10 & 0.85 & 0.84 & 0.85 \\
15 & 0.83 & 0.86 & 0.83 \\
\hline
\end{tabular}
\caption{$J=30$, $G=5$, $n=500$}
\end{subtable}
\hfill
\begin{subtable}[t]{0.45\textwidth}
\centering
\label{tab:Sen EMC4}
\begin{tabular}{|c|c c c|}
\hline
\diagbox{$c_{\sigma}$}{$c_{\theta}$} & 0.25 & 0.5 & 0.75 \\ 
\hline
5 & 1.00 & 1.00 & 1.00 \\
10 & 1.00 & 1.00 & 1.00 \\
15 & 1.00 & 1.00 & 1.00 \\
\hline
\end{tabular}
\caption{$J=30$, $G=5$, $n=2000$}
\end{subtable}
\label{tab:SenEMC}
\end{table}

\begin{table}[htbp]
\centering
\caption{Sensitivity Analysis of ACC.}
\begin{subtable}[t]{0.45\textwidth}
\centering
\label{tab:Sen ACC1}
\begin{tabular}{|c|c c c|}
\hline
\diagbox{$c_{\sigma}$}{$c_{\theta}$} & 0.25 & 0.5 & 0.75 \\ 
\hline
5 & 1.000 & 1.000 & 1.000 \\
10 & 1.000 & 1.000 & 1.000 \\
15 & 1.000 & 1.000 & 1.000 \\
\hline
\end{tabular}
\caption{$J=15$, $G=3$, $n=500$}
\end{subtable}
\hfill
\begin{subtable}[t]{0.45\textwidth}
\centering
\label{tab:Sen ACC2}
\begin{tabular}{|c|c c c|}
\hline
\diagbox{$c_{\sigma}$}{$c_{\theta}$} & 0.25 & 0.5 & 0.75 \\ 
\hline
5 & 1.000 & 1.000 & 1.000 \\
10 & 1.000 & 1.000 & 1.000 \\
15 & 1.000 & 1.000 & 1.000 \\
\hline
\end{tabular}
\caption{$J=15$, $G=3$, $n=2000$}
\end{subtable}
\hfill
\begin{subtable}[t]{0.45\textwidth}
\centering
\label{tab:Sen ACC3}
\begin{tabular}{|c|c c c|}
\hline
\diagbox{$c_{\sigma}$}{$c_{\theta}$} & 0.25 & 0.5 & 0.75 \\ 
\hline
5 & 0.998 & 0.998 & 0.998 \\
10 & 0.998 & 0.997 & 0.998 \\
15 & 0.997 & 0.998 & 0.997 \\
\hline
\end{tabular}
\caption{$J=30$, $G=5$, $n=500$}
\end{subtable}
\hfill
\begin{subtable}[t]{0.45\textwidth}
\centering
\label{tab:Sen ACC4}
\begin{tabular}{|c|c c c|}
\hline
\diagbox{$c_{\sigma}$}{$c_{\theta}$} & 0.25 & 0.5 & 0.75 \\ 
\hline
5 & 1.000 & 1.000 & 1.000 \\
10 & 1.000 & 1.000 & 1.000 \\
15 & 1.000 & 1.000 & 1.000 \\
\hline
\end{tabular}
\caption{$J=30$, $G=5$, $n=2000$}
\end{subtable}
\label{tab:SenACC}
\end{table}

\begin{table}[htbp]
\centering
\caption{Sensitivity Analysis of Computation time(s).}
\begin{subtable}[t]{0.45\textwidth}
\centering
\label{tab:Sen Time1}
\begin{tabular}{|c|c c c|}
\hline
\diagbox{$c_{\sigma}$}{$c_{\theta}$} & 0.25 & 0.5 & 0.75 \\ 
\hline
5 & 0.13 & 0.13 & 0.13 \\
10 & 0.13 & 0.13 & 0.12 \\
15 & 0.10 & 0.09 & 0.08 \\
\hline
\end{tabular}
\caption{$J=15$, $G=3$, $n=500$}
\end{subtable}
\hfill
\begin{subtable}[t]{0.45\textwidth}
\centering
\label{tab:Sen Time2}
\begin{tabular}{|c|c c c|}
\hline
\diagbox{$c_{\sigma}$}{$c_{\theta}$} & 0.25 & 0.5 & 0.75 \\ 
\hline
5 & 0.10 & 0.09 & 0.10 \\
10 & 0.10 & 0.09 & 0.09 \\
15 & 0.07 & 0.06 & 0.06 \\
\hline
\end{tabular}
\caption{$J=15$, $G=3$, $n=2000$}
\end{subtable}
\hfill
\begin{subtable}[t]{0.45\textwidth}
\centering
\label{tab:Sen Time3}
\begin{tabular}{|c|c c c|}
\hline
\diagbox{$c_{\sigma}$}{$c_{\theta}$} & 0.25 & 0.5 & 0.75 \\ 
\hline
5 & 0.52 & 0.51 & 0.50 \\
10 & 0.49 & 0.47 & 0.41 \\
15 & 0.36 & 0.33 & 0.30 \\
\hline
\end{tabular}
\caption{$J=30$, $G=5$, $n=500$}
\end{subtable}
\hfill
\begin{subtable}[t]{0.45\textwidth}
\centering
\label{tab:Sen Time4}
\begin{tabular}{|c|c c c|}
\hline
\diagbox{$c_{\sigma}$}{$c_{\theta}$} & 0.25 & 0.5 & 0.75 \\ 
\hline
5 & 0.38 & 0.37 & 0.37 \\
10 & 0.37 & 0.33 & 0.28 \\
15 & 0.23 & 0.22 & 0.24 \\
\hline
\end{tabular}
\caption{$J=30$, $G=5$, $n=2000$}
\end{subtable}
\label{tab:SenTime}
\end{table}

\section{Extension to Hierarchical Factor Analysis}\label{appen:hfm}

\subsection{Constrained Optimisation for Exploratory Hierarchical Factor Analysis}

To further demonstrate the advantages of the constraint-based approach, we discuss how it can be extended for exploratory hierarchical factor analysis. Following the terminology adopted in \cite{yung1999relationship}, 
we consider general
hierarchical factor models.
Such a model has several layers of factors. In each layer, each observed variable loads on exactly one of the factors in that layer. The
numbering of the layers is determined by the number of factors in the layer, starting from the layer with the largest number of factors. Each factor in a lower layer is nested within a factor in a higher layer, in the sense that the variables loading on the lower-layer factor must also all load on a higher-layer factor. All the factors are assumed to be uncorrelated (i.e., $\Phi$ is an identity matrix),  though this assumption may be relaxed to allow some correlations between factors within the same layer as in the extended bi-factor model. 
\begin{figure}
    \centering
    \begin{subfigure}[b]{0.6\textwidth}
        \centering
        \begin{tikzpicture}[x=0.75pt,y=0.75pt,yscale=-1,xscale=1]

\draw   (120,129.29) -- (142,129.29) -- (142,150.71) -- (120,150.71) -- cycle ;
\draw   (200,129.29) -- (222,129.29) -- (222,150.71) -- (200,150.71) -- cycle ;
\draw   (249,129.29) -- (271,129.29) -- (271,150.71) -- (249,150.71) -- cycle ;
\draw   (38,129.29) -- (60,129.29) -- (60,150.71) -- (38,150.71) -- cycle ;
\draw   (279,129.29) -- (301,129.29) -- (301,150.71) -- (279,150.71) -- cycle ;
\draw   (169,129.29) -- (191,129.29) -- (191,150.71) -- (169,150.71) -- cycle ;
\draw   (89,129.29) -- (111,129.29) -- (111,150.71) -- (89,150.71) -- cycle ;
\draw   (330,129.29) -- (352,129.29) -- (352,150.71) -- (330,150.71) -- cycle ;
\draw   (142.29,191) .. controls (142.29,182.87) and (148.87,176.29) .. (157,176.29) .. controls (165.13,176.29) and (171.71,182.87) .. (171.71,191) .. controls (171.71,199.13) and (165.13,205.71) .. (157,205.71) .. controls (148.87,205.71) and (142.29,199.13) .. (142.29,191) -- cycle ;
\draw   (181.57,91) .. controls (181.57,82.87) and (188.16,76.29) .. (196.29,76.29) .. controls (204.41,76.29) and (211,82.87) .. (211,91) .. controls (211,99.13) and (204.41,105.71) .. (196.29,105.71) .. controls (188.16,105.71) and (181.57,99.13) .. (181.57,91) -- cycle ;
\draw   (220.29,191) .. controls (220.29,182.87) and (226.87,176.29) .. (235,176.29) .. controls (243.13,176.29) and (249.71,182.87) .. (249.71,191) .. controls (249.71,199.13) and (243.13,205.71) .. (235,205.71) .. controls (226.87,205.71) and (220.29,199.13) .. (220.29,191) -- cycle ;
\draw   (101.57,91) .. controls (101.57,82.87) and (108.16,76.29) .. (116.29,76.29) .. controls (124.41,76.29) and (131,82.87) .. (131,91) .. controls (131,99.13) and (124.41,105.71) .. (116.29,105.71) .. controls (108.16,105.71) and (101.57,99.13) .. (101.57,91) -- cycle ;
\draw   (300.29,191) .. controls (300.29,182.87) and (306.87,176.29) .. (315,176.29) .. controls (323.13,176.29) and (329.71,182.87) .. (329.71,191) .. controls (329.71,199.13) and (323.13,205.71) .. (315,205.71) .. controls (306.87,205.71) and (300.29,199.13) .. (300.29,191) -- cycle ;
\draw   (61.29,191) .. controls (61.29,182.87) and (67.87,176.29) .. (76,176.29) .. controls (84.13,176.29) and (90.71,182.87) .. (90.71,191) .. controls (90.71,199.13) and (84.13,205.71) .. (76,205.71) .. controls (67.87,205.71) and (61.29,199.13) .. (61.29,191) -- cycle ;
\draw   (260.57,91) .. controls (260.57,82.87) and (267.16,76.29) .. (275.29,76.29) .. controls (283.41,76.29) and (290,82.87) .. (290,91) .. controls (290,99.13) and (283.41,105.71) .. (275.29,105.71) .. controls (267.16,105.71) and (260.57,99.13) .. (260.57,91) -- cycle ;
\draw    (100,105.71) -- (50.82,128.45) ;
\draw [shift={(49,129.29)}, rotate = 335.19] [color={rgb, 255:red, 0; green, 0; blue, 0 }  ][line width=0.75]    (10.93,-3.29) .. controls (6.95,-1.4) and (3.31,-0.3) .. (0,0) .. controls (3.31,0.3) and (6.95,1.4) .. (10.93,3.29)   ;
\draw    (100,105.71) -- (100,125) ;
\draw [shift={(100,127)}, rotate = 270] [color={rgb, 255:red, 0; green, 0; blue, 0 }  ][line width=0.75]    (10.93,-3.29) .. controls (6.95,-1.4) and (3.31,-0.3) .. (0,0) .. controls (3.31,0.3) and (6.95,1.4) .. (10.93,3.29)   ;
\draw    (131,105.71) -- (131,127.29) ;
\draw [shift={(131,129.29)}, rotate = 270] [color={rgb, 255:red, 0; green, 0; blue, 0 }  ][line width=0.75]    (10.93,-3.29) .. controls (6.95,-1.4) and (3.31,-0.3) .. (0,0) .. controls (3.31,0.3) and (6.95,1.4) .. (10.93,3.29)   ;
\draw    (131,105.71) -- (178.2,128.42) ;
\draw [shift={(180,129.29)}, rotate = 205.69] [color={rgb, 255:red, 0; green, 0; blue, 0 }  ][line width=0.75]    (10.93,-3.29) .. controls (6.95,-1.4) and (3.31,-0.3) .. (0,0) .. controls (3.31,0.3) and (6.95,1.4) .. (10.93,3.29)   ;
\draw    (180,105.71) -- (50.97,128.93) ;
\draw [shift={(49,129.29)}, rotate = 349.8] [color={rgb, 255:red, 0; green, 0; blue, 0 }  ][line width=0.75]    (10.93,-3.29) .. controls (6.95,-1.4) and (3.31,-0.3) .. (0,0) .. controls (3.31,0.3) and (6.95,1.4) .. (10.93,3.29)   ;
\draw    (180,105.71) -- (101.93,126.49) ;
\draw [shift={(100,127)}, rotate = 345.1] [color={rgb, 255:red, 0; green, 0; blue, 0 }  ][line width=0.75]    (10.93,-3.29) .. controls (6.95,-1.4) and (3.31,-0.3) .. (0,0) .. controls (3.31,0.3) and (6.95,1.4) .. (10.93,3.29)   ;
\draw    (180,105.71) -- (132.8,128.42) ;
\draw [shift={(131,129.29)}, rotate = 334.31] [color={rgb, 255:red, 0; green, 0; blue, 0 }  ][line width=0.75]    (10.93,-3.29) .. controls (6.95,-1.4) and (3.31,-0.3) .. (0,0) .. controls (3.31,0.3) and (6.95,1.4) .. (10.93,3.29)   ;
\draw    (180,105.71) -- (180,127.29) ;
\draw [shift={(180,129.29)}, rotate = 270] [color={rgb, 255:red, 0; green, 0; blue, 0 }  ][line width=0.75]    (10.93,-3.29) .. controls (6.95,-1.4) and (3.31,-0.3) .. (0,0) .. controls (3.31,0.3) and (6.95,1.4) .. (10.93,3.29)   ;
\draw    (211,105.71) -- (211,127.29) ;
\draw [shift={(211,129.29)}, rotate = 270] [color={rgb, 255:red, 0; green, 0; blue, 0 }  ][line width=0.75]    (10.93,-3.29) .. controls (6.95,-1.4) and (3.31,-0.3) .. (0,0) .. controls (3.31,0.3) and (6.95,1.4) .. (10.93,3.29)   ;
\draw    (211,105.71) -- (258.2,128.42) ;
\draw [shift={(260,129.29)}, rotate = 205.69] [color={rgb, 255:red, 0; green, 0; blue, 0 }  ][line width=0.75]    (10.93,-3.29) .. controls (6.95,-1.4) and (3.31,-0.3) .. (0,0) .. controls (3.31,0.3) and (6.95,1.4) .. (10.93,3.29)   ;
\draw    (211,105.71) -- (288.08,128.71) ;
\draw [shift={(290,129.29)}, rotate = 196.61] [color={rgb, 255:red, 0; green, 0; blue, 0 }  ][line width=0.75]    (10.93,-3.29) .. controls (6.95,-1.4) and (3.31,-0.3) .. (0,0) .. controls (3.31,0.3) and (6.95,1.4) .. (10.93,3.29)   ;
\draw    (211,105.71) -- (339.03,128.93) ;
\draw [shift={(341,129.29)}, rotate = 190.28] [color={rgb, 255:red, 0; green, 0; blue, 0 }  ][line width=0.75]    (10.93,-3.29) .. controls (6.95,-1.4) and (3.31,-0.3) .. (0,0) .. controls (3.31,0.3) and (6.95,1.4) .. (10.93,3.29)   ;
\draw    (260,105.71) -- (212.8,128.42) ;
\draw [shift={(211,129.29)}, rotate = 334.31] [color={rgb, 255:red, 0; green, 0; blue, 0 }  ][line width=0.75]    (10.93,-3.29) .. controls (6.95,-1.4) and (3.31,-0.3) .. (0,0) .. controls (3.31,0.3) and (6.95,1.4) .. (10.93,3.29)   ;
\draw    (260,105.71) -- (260,127.29) ;
\draw [shift={(260,129.29)}, rotate = 270] [color={rgb, 255:red, 0; green, 0; blue, 0 }  ][line width=0.75]    (10.93,-3.29) .. controls (6.95,-1.4) and (3.31,-0.3) .. (0,0) .. controls (3.31,0.3) and (6.95,1.4) .. (10.93,3.29)   ;
\draw    (290,105.71) -- (290,127.29) ;
\draw [shift={(290,129.29)}, rotate = 270] [color={rgb, 255:red, 0; green, 0; blue, 0 }  ][line width=0.75]    (10.93,-3.29) .. controls (6.95,-1.4) and (3.31,-0.3) .. (0,0) .. controls (3.31,0.3) and (6.95,1.4) .. (10.93,3.29)   ;
\draw    (290,105.71) -- (339.18,128.45) ;
\draw [shift={(341,129.29)}, rotate = 204.81] [color={rgb, 255:red, 0; green, 0; blue, 0 }  ][line width=0.75]    (10.93,-3.29) .. controls (6.95,-1.4) and (3.31,-0.3) .. (0,0) .. controls (3.31,0.3) and (6.95,1.4) .. (10.93,3.29)   ;
\draw    (60.43,176.29) -- (49.82,152.54) ;
\draw [shift={(49,150.71)}, rotate = 65.92] [color={rgb, 255:red, 0; green, 0; blue, 0 }  ][line width=0.75]    (10.93,-3.29) .. controls (6.95,-1.4) and (3.31,-0.3) .. (0,0) .. controls (3.31,0.3) and (6.95,1.4) .. (10.93,3.29)   ;
\draw    (90.43,176.29) -- (99.3,152.59) ;
\draw [shift={(100,150.71)}, rotate = 110.52] [color={rgb, 255:red, 0; green, 0; blue, 0 }  ][line width=0.75]    (10.93,-3.29) .. controls (6.95,-1.4) and (3.31,-0.3) .. (0,0) .. controls (3.31,0.3) and (6.95,1.4) .. (10.93,3.29)   ;
\draw    (141.43,176.29) -- (131.76,152.57) ;
\draw [shift={(131,150.71)}, rotate = 67.81] [color={rgb, 255:red, 0; green, 0; blue, 0 }  ][line width=0.75]    (10.93,-3.29) .. controls (6.95,-1.4) and (3.31,-0.3) .. (0,0) .. controls (3.31,0.3) and (6.95,1.4) .. (10.93,3.29)   ;
\draw    (171.43,176.29) -- (179.36,152.61) ;
\draw [shift={(180,150.71)}, rotate = 108.53] [color={rgb, 255:red, 0; green, 0; blue, 0 }  ][line width=0.75]    (10.93,-3.29) .. controls (6.95,-1.4) and (3.31,-0.3) .. (0,0) .. controls (3.31,0.3) and (6.95,1.4) .. (10.93,3.29)   ;
\draw    (221.43,176.29) -- (211.76,152.57) ;
\draw [shift={(211,150.71)}, rotate = 67.81] [color={rgb, 255:red, 0; green, 0; blue, 0 }  ][line width=0.75]    (10.93,-3.29) .. controls (6.95,-1.4) and (3.31,-0.3) .. (0,0) .. controls (3.31,0.3) and (6.95,1.4) .. (10.93,3.29)   ;
\draw    (248.43,176.29) -- (259.18,152.54) ;
\draw [shift={(260,150.71)}, rotate = 114.35] [color={rgb, 255:red, 0; green, 0; blue, 0 }  ][line width=0.75]    (10.93,-3.29) .. controls (6.95,-1.4) and (3.31,-0.3) .. (0,0) .. controls (3.31,0.3) and (6.95,1.4) .. (10.93,3.29)   ;
\draw    (301.43,176.29) -- (290.82,152.54) ;
\draw [shift={(290,150.71)}, rotate = 65.92] [color={rgb, 255:red, 0; green, 0; blue, 0 }  ][line width=0.75]    (10.93,-3.29) .. controls (6.95,-1.4) and (3.31,-0.3) .. (0,0) .. controls (3.31,0.3) and (6.95,1.4) .. (10.93,3.29)   ;
\draw    (329.43,176.29) -- (340.18,152.54) ;
\draw [shift={(341,150.71)}, rotate = 114.35] [color={rgb, 255:red, 0; green, 0; blue, 0 }  ][line width=0.75]    (10.93,-3.29) .. controls (6.95,-1.4) and (3.31,-0.3) .. (0,0) .. controls (3.31,0.3) and (6.95,1.4) .. (10.93,3.29)   ;

\draw (44,132) node [anchor=north west][inner sep=0.75pt]   [align=left] {1};
\draw (95,132) node [anchor=north west][inner sep=0.75pt]   [align=left] {5};
\draw (126,132) node [anchor=north west][inner sep=0.75pt]   [align=left] {6};
\draw (171,132) node [anchor=north west][inner sep=0.75pt]   [align=left] {10};
\draw (202.5,132) node [anchor=north west][inner sep=0.75pt]   [align=left] {11};
\draw (251,132) node [anchor=north west][inner sep=0.75pt]   [align=left] {15};
\draw (281,132) node [anchor=north west][inner sep=0.75pt]   [align=left] {16};
\draw (332,132) node [anchor=north west][inner sep=0.75pt]   [align=left] {20};
\draw (187,82) node [anchor=north west][inner sep=0.75pt]   [align=left] {$\displaystyle F_{1}$};
\draw (107,82) node [anchor=north west][inner sep=0.75pt]   [align=left] {$\displaystyle F_{2}$};
\draw (266,82) node [anchor=north west][inner sep=0.75pt]   [align=left] {$\displaystyle F_{3}$};
\draw (67,182) node [anchor=north west][inner sep=0.75pt]   [align=left] {$\displaystyle F_{4}$};
\draw (148,182) node [anchor=north west][inner sep=0.75pt]   [align=left] {$\displaystyle F_{5}$};
\draw (226,182) node [anchor=north west][inner sep=0.75pt]   [align=left] {$\displaystyle F_{6}$};
\draw (306,182) node [anchor=north west][inner sep=0.75pt]   [align=left] {$\displaystyle F_{7}$};
\draw (64.21,136.79) node [anchor=north west][inner sep=0.75pt]   [align=left] {$\displaystyle \cdots $};
\draw (223.21,136.79) node [anchor=north west][inner sep=0.75pt]   [align=left] {$\displaystyle \cdots $};
\draw (144.71,136.79) node [anchor=north west][inner sep=0.75pt]   [align=left] {$\displaystyle \cdots $};
\draw (304.21,136.79) node [anchor=north west][inner sep=0.75pt]   [align=left] {$\displaystyle \cdots $};

\end{tikzpicture}
        \caption{The path diagram of a three-layer hierarchical factor model.}
        \label{fig: example path}
    \end{subfigure}
    \hfill
    \hfill
    \begin{subfigure}[b]{0.5\textwidth}
        \centering     

\begin{tikzpicture}[x=0.75pt,y=0.75pt,yscale=-1,xscale=1]

\draw   (149,20) -- (171,20) -- (171,41.43) -- (149,41.43) -- cycle ;
\draw    (160,41.43) -- (160,51) ;
\draw   (215,100) -- (237,100) -- (237,121.43) -- (215,121.43) -- cycle ;
\draw   (104,60) -- (126,60) -- (126,81.43) -- (104,81.43) -- cycle ;
\draw   (84,100) -- (106,100) -- (106,121.43) -- (84,121.43) -- cycle ;
\draw   (172,100) -- (194,100) -- (194,121.43) -- (172,121.43) -- cycle ;
\draw   (128,100) -- (150,100) -- (150,121.43) -- (128,121.43) -- cycle ;
\draw   (195,60) -- (217,60) -- (217,81.43) -- (195,81.43) -- cycle ;
\draw    (115,51) -- (206,51) ;
\draw    (115,51) -- (115,60) ;
\draw    (206,51) -- (206,60) ;
\draw    (115,81.43) -- (115,91) ;
\draw    (206,81.43) -- (206,91) ;
\draw    (95,91) -- (139,91) ;
\draw    (183,91) -- (226,91) ;
\draw    (95,91) -- (95,100) ;
\draw    (139,91) -- (139,100) ;
\draw    (183,91) -- (183,100) ;
\draw    (226,91) -- (226,100) ;

\draw (151,22) node [anchor=north west][inner sep=0.75pt]   [align=left] {$\displaystyle F_{1}$};
\draw (106,64) node [anchor=north west][inner sep=0.75pt]   [align=left] {$\displaystyle F_{2}$};
\draw (197,64) node [anchor=north west][inner sep=0.75pt]   [align=left] {$\displaystyle F_{3}$};
\draw (86,104) node [anchor=north west][inner sep=0.75pt]   [align=left] {$\displaystyle F_{4}$};
\draw (130,104) node [anchor=north west][inner sep=0.75pt]   [align=left] {$\displaystyle F_{5}$};
\draw (174,103) node [anchor=north west][inner sep=0.75pt]   [align=left] {$\displaystyle F_{6}$};
\draw (217,103) node [anchor=north west][inner sep=0.75pt]   [align=left] {$\displaystyle F_{7}$};

\end{tikzpicture}

        \caption{The corresponding factor
hierarchy.}
        \label{fig: example hierarchy}
    \end{subfigure}
    
    \caption{The illustrative example of a three-layer hierarchical factor model.}
    \label{fig:hier}
\end{figure}
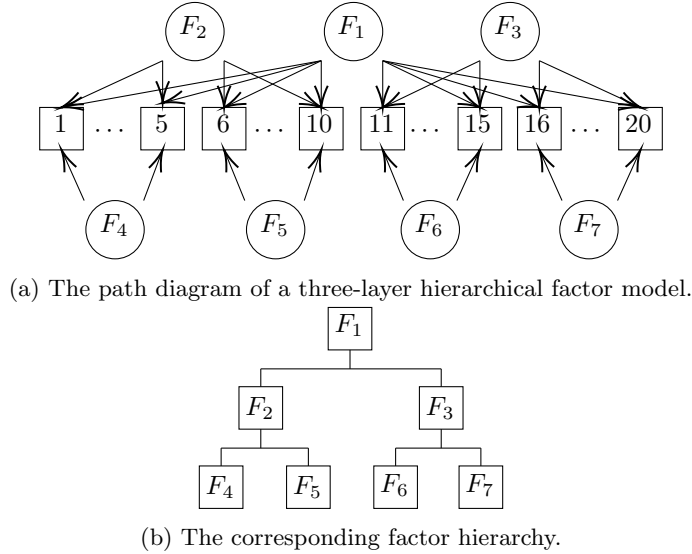

Panel (a) of Figure \ref{fig:hier} provides the path diagram of a hierarchical factor model that has three layers, with factor $F_1$ in layer 3, factors $F_2$ and $F_3$ in layer 2, and factors $F_4$-$F_7$ in layer 1. The corresponding factor hierarchy is summarised in Panel (b) of 
Figure \ref{fig:hier} that takes the form of a tree, where $F_2$ and $F_3$ are nested within $F_1$, $F_4$ and $F_5$ are nested within $F_2$, and $F_6$ and $F_7$ are nested within $F_3$. 
In what follows, we show how the loading structure of this three-layer hierarchical model can be learned by a constrained optimisation method, assuming that the factor hierarchy in Panel (b) of Figure \ref{fig:hier} is known while the variables loading on each factor are unknown. The goal is to learn how the observed variables load on the seven factors.

Following the same notation for bi-factor analysis, the population covariance matrix of observed variables under the hierarchical factor model can be written as 
$$\Sigma = \Lambda \Lambda^\top + \Psi,$$
where $\Lambda$ is a $J\times 7$ matrix, and $\Psi$ is a $J\times J$ diagonal matrix. Note that we no longer need the correlation matrix $\Phi$ in the expression as it is now an identity matrix. The constraints implied by the hierarchical factor structure become: 

\begin{equation}\label{eq:constraint2}
\begin{aligned}
   & \lambda_{j2}\lambda_{j3} = 0,   ~~\lambda_{j2}\lambda_{j6} = 0,     ~~ \lambda_{j2}\lambda_{j7} = 0,     \\
   & \lambda_{j3}\lambda_{j4} = 0, ~~\lambda_{j3}\lambda_{j5} = 0,    \\ 
   & \lambda_{j4}\lambda_{j5} = 0,  ~~   \lambda_{j6}\lambda_{j7} = 0,   ~~ j=1, ..., J.\\ 
\end{aligned}
\end{equation}
Consequently, the corresponding hierarchical factor model can be learned by minimising the loss function $l( \Lambda \Lambda^\top + \Psi(\boldsymbol{\psi}); S)$, subject to the constraints in \eqref{eq:constraint2}. 

Although the above discussion focuses on the specific hierarchical factor structure in Figure~\ref{fig:hier}, 
when given a different factor hierarchy,
it is easy to derive similar constraints as in \eqref{eq:constraint2} by induction. Based on the constraints,  the corresponding hierarchical factor model can be learned by an ALM. 

Finally, we note that the factor hierarchy is typically unknown in practice. In that case, we need an algorithm that simultaneously learns the factor hierarchy and the variable loadings on the hierarchical factors. As there are exponentially many choices for the structure of factor hierarchy,  this problem is more challenging than the setting when the factor hierarchy is known.  It is also more challenging than exploratory bi-factor analysis with unknown group factors, as the bi-factor model has a simple two-layer factor hierarchy that is completely determined by the number of factors. 

\subsection{Simulation}
In this section, we examine the recovery of the hierarchical structure of our method. For $J\in \{20,40\}$ and $N\in\{500,2000\}$, a data generation model is considered, resulting in a total of 4 simulation settings. With slight abuse of notation, we denote by $\mathcal{B}_{g}^{*}$ as the true item groups related to the $g$th factor. In the data generation model, $B_{1}^{*} = \{1,\ldots,J\}$, $B_{2}^{*} = \{1,\ldots,J/2\}$, $B_{3}^{*} = \{J/2,\ldots,J\}$, $B_{4}^{*} = \{1,\ldots,J/4\}$, $B_{5}^{*} = \{J/4,\ldots,J/2\}$, $B_{6}^{*} = \{J/2,\ldots,3J/4\}$, $B_{7}^{*} = \{3J/4,\ldots,J\}$. $\Psi^* = \mathbb{I}_{J\times J}$, and $\Lambda^*$ follows
\begin{equation}
\begin{aligned}
\lambda_{jk}^{*} = \left\{
    \begin{array}{l}
        u_{jk} \mbox{~~if~~} k=1; \\
        0 \mbox{~~if~~} k>1 ,j \notin \mathcal{B}_{k-1}^{*}; \\
        (1-2x_{jk})u_{jk} \mbox{~~if~~} k>1, j\in\mathcal{B}_{k-1}^{*},
    \end{array}
\right.
\end{aligned}
\end{equation}
for $j = 1,\ldots,J$ and $k = 1,\ldots, G+1$. Here, $u_{jk}$s are i.i.d., following a Uniform$(0.2,1)$ distribution, and $x_{jk}$s are i.i.d., following a Bernoulli$(0.5)$ distribution.

The estimated parameters $\hat{\Lambda}$ and $\hat{\Psi}$ follow the same ALM algorithm in Section~\ref{subsec:alm} except that the distance between the estimate and the space of the hierarchical factor loading matrices measured by $$\max_{j \in \{1, ..., J\}} \tilde{h}(|\lambda_{j2}^{(t)}|, ..., |\lambda_{j,G+1}^{(t)}|), $$
where the function $\tilde{h}$ returns the third-largest value of a vector. The estimated hierarchical factor model structure is then given by $$\hat{\mathcal B}_g = \{j: |\lambda_{j,g+1}^{(T)}| > \delta_2\}.$$
We also choose $\delta_{2} = 10^{-2}$ on the following simulation study.

Since label-switching problem exists in factors that are nested within the same hierarchical factor, there exists 8 possible permutations of labels resulting in the same hierarchical structure. We denote by $\mathcal{R}$ as the set of the 8 permutations. Then, the evaluation criteria for the recovery of the hierarchical structure are defined as:
\begin{itemize}
    \item Exact Match Criterion(EMC): $\max_{\sigma\in \mathcal{R}}\prod_{g=1}^{G}\mathbf{1}(\mathcal{B}_{\sigma(g)} = \mathcal{B}_{g}^{*})$, which equals 1 when the bi-factor structure is correctly learned and 0 otherwise.
    \item Average Correctness Criterion(ACC): $\max_{\sigma\in\mathcal{R}}\sum_{g=1}^{G}(\vert\mathcal{B}_{g}^{*} \cap \mathcal{B}_{\sigma(g)}\vert + \vert\mathcal{B}_{\sigma(g)}^{C} \cap \mathcal{B}_{g}^{*C}\vert) / (JG)$.
\end{itemize}

For each setting, we first generate $\Lambda^{*}$ once and use them to generate 100 datasets. The averaged results under 100 replication are shown in Table~\ref{tab:hierarchy}. From the simulation results, we find that our method performs well on the recovery of hierarchical structure.

\begin{table}[h]
    \centering
    \caption{Simulation results of the recovery of hierarchical factor structure.}
    \label{tab:hierarchy}
    \begin{tabular}{ |c | c | c | c| }
    \hline
    $J$ & $N$ & EMC & ACC \\
    \hline
    20 & 500 & 0.94 & 0.998 \\
    \hline
     & 2000 & 1.00 & 1.000\\
    \hline
    40 & 500 & 0.88 & 0.988\\
    \hline
     & 2000 & 1.00 & 1.000\\
    \hline

    \end{tabular}
\end{table}

\section{Extraversion Scale Item Key}\label{appen:ItemKey}

\begin{table}[H]
    \centering
    \centering
    \caption{Extraversion Item Key}
    \label{tab:ItemKey}
    \begin{tabular}{| c | l |  l | l |}
    \hline
    Item & Sign & Facet & Item\\
    \hline
    1 & +E1 & Friendliness & Make friends easily.\\
    2 & +E1 & Friendliness & Feel comfortable around people.\\
    3 & $-$E1 & Friendliness & Avoid contacts with others.\\
    4 & $-$E1 & Friendliness & Keep others at a distance. \\
     & & &  \\
    5 & +E2 & Gregariousness & Love large parties.\\
    6 & +E2 & Gregariousness & Talk to a lot of different people at parties.\\
    7 & $-$E2 & Gregariousness & Prefer to be alone. \\
    8 & $-$E2 & Gregariousness & Avoid crowds. \\
    & & &  \\
    9 & +E3 & Assertiveness & Take charge. \\
    10 & +E3 & Assertiveness & Try to lead others. \\
    11 & +E3 & Assertiveness & Take control of things. \\
    12 & $-$E3 & Assertiveness & Wait for others to lead the way.\\
    & & &  \\
    13 & +E4 & Activity Level & Am always busy.\\
    14 & +E4 & Activity Level & Am always on the go.\\
    15 & +E4 & Activity Level & Do a lot in my spare time.\\
    16 & $-$E4 & Activity Level & Like to take it easy.\\
    & & &  \\
    17 & +E5 & Excitement-Seeking & Love excitement.\\
    18 & +E5 & Excitement-Seeking & Seek adventure.\\
    19 & +E5 & Excitement-Seeking & Enjoy being reckless.\\
    20  & +E5 & Excitement-Seeking & Act wild and crazy.\\
    & & &  \\
    21 & +E6 & Cheerfulness & Radiate joy.\\
    22 & +E6 & Cheerfulness & Have a lot of fun.\\
    23 & +E6 & Cheerfulness & Love life.\\
    24 & +E6 & Cheerfulness & Look at the bright side of life.\\
    \hline
     \end{tabular}
\end{table}

\section{Real Data Analysis using Bi-factor Rotation Method}\label{appen:real data rotation}
In this section, we present the results of the same data in Section~\ref{sec:real data} by bi-factor rotation method as a comparison with our proposed method. 

Using a candidate set $\mathcal{G} = \{2,\ldots,12\}$, the BIC procedure of exploratory factor analysis given in Section~\ref{subsec:study 2} selects eight factors in total, which coincide with the number of factors selected by the BIC procedure of our proposed method. By applying the bi-factor rotation method\citep{jennrich2012exploratory}, we get the rotation solutions $\hat{\Lambda}^{oblq}$ in Table~\ref{tab:bfrt-loading} and $\hat{\Phi}^{oblq}$ in equation~\eqref{eq:bfrt-phi}.
\begin{table}[h]
    \centering
    \caption{Estimated loading matrix $\hat{\Lambda}^{oblq}$ with seven group factors. }
    \label{tab:bfrt-loading}
    \begin{tabular}{| r | r |r | r | r | r | r | r | r | r |}
    \hline
    Items & Sign & General & G1 & G2 & G3 & G4
    & G5 & G6 & G7\\
    \hline
        1& +E1 &0.86 & 0.01 & -0.06 & -0.08 & -0.03 & -0.04 & 0.42 & -0.08 \\ 
        2& +E1 &0.85 & 0.03 & -0.11 & 0.03 & 0.06 & -0.12 & 0.07 & -0.01 \\
        3& $-$E1&0.91 & -0.02 & -0.12 & -0.01 & 0.03 & -0.09 & -0.02 & -0.05  \\ 
        4& $-$E1&0.87 & -0.14 & -0.04 & -0.01 & -0.10 & -0.14 & -0.03 & -0.20 \\         
        5& +E2&0.88 & 0.69 & 0.00 & 0.00 & -0.02 & 0.00 & -0.01 & 0.01 \\ 
        6& +E2&0.92 & 0.25 & 0.03 & -0.12 & 0.09 & 0.05 & 0.22 & -0.03 \\ 
        7& $-$E2&0.72 & -0.06 & -0.03 & -0.04 & -0.08 & -0.16 & -0.21 & -0.12 \\ 
        8& $-$E2&0.85 & 0.22 & -0.02 & -0.05 & -0.06 & -0.07 & -0.29 & -0.08\\ 
        9& +E3&0.52 & 0.02 & 0.02 & 0.00 & 0.79 & 0.02 & 0.04 & -0.03  \\ 
        10& +E3&0.52 & 0.04 & 0.00 & 0.00 & 0.75 & -0.03 & 0.03 & 0.01 \\ 
        11& +E3& 0.44 & -0.03 & 0.00 & 0.05 & 0.62 & 0.04 & -0.08 & 0.01 \\
        12& $-$E3&0.55 & -0.09 & -0.04 & -0.02 & 0.62 & -0.05 & -0.06 & 0.06 \\ 
        13& +E4& 0.32 & 0.05 & 0.01 & 0.00 & 0.02 & 0.82 & -0.02 & -0.06 \\ 
        14& +E4& 0.51 & -0.07 & 0.02 & -0.02 & -0.01 & 0.74 & 0.06 & 0.06  \\ 
        15& +E4&0.49 & 0.02 & -0.08 & 0.15 & -0.02 & 0.51 & -0.06 & 0.14\\ 
        16& $-$E4&0.19 & -0.14 & -0.04 & -0.14 & 0.08 & 0.37 & -0.19 & -0.07 \\
        17& +E5& 0.46 & 0.09 & -0.03 & -0.04 & 0.02 & 0.02 & 0.06 & 0.49  \\ 
        18& +E5& 0.53 & -0.05 & 0.02 & 0.00 & 0.01 & 0.03 & -0.04 & 0.62\\ 
        19& +E5& 0.28 & 0.05 & 0.48 & -0.02 & 0.02 & -0.12 & 0.00 & 0.33 \\ 
        20& +E5& 0.48 & 0.00 & 1.10 & 0.00 & 0.00 & 0.01 & 0.00 & -0.02 \\ 
        21& +E6& 0.64 & -0.12 & 0.04 & 0.26 & -0.04 & 0.01 & 0.28 & -0.01  \\ 
        22& +E6&0.69 & 0.06 & 0.11 & 0.40 & -0.01 & 0.01 & 0.09 & 0.08 \\ 
        23& +E6&0.63 & 0.02 & -0.01 & 0.63 & 0.03 & 0.01 & -0.05 & -0.01  \\ 
        24& +E6& 0.58 & -0.04 & -0.03 & 0.60 & 0.00 & -0.02 & 0.01 & -0.04 \\ \hline
    \end{tabular}
\end{table}

\begin{equation}\label{eq:bfrt-phi}
\begin{aligned}
\hat{\Phi}^{oblq} = \left(
\begin{matrix}
1 & 0& 0 & 0& 0 & 0 & 0 & 0 \\
0 & 1 & 0.19 & -0.14 & -0.17 & -0.15 & 0.10 & 0.05 \\ 
0 & 0.19 & 1  & -0.10 & 0.00 & -0.07 & 0.11 & 0.39 \\ 
0 & -0.14 & -0.10 & 1 & 0.02 & 0.07 & 0.06 & 0.07 \\ 
0 & -0.17 & 0.00 & 0.02 & 1 & 0.21 &-0.07 & 0.11 \\ 
0 & -0.15 & -0.07 & 0.07 & 0.21 & 1 & -0.09 & 0.01 \\ 
0 & 0.10 & 0.11 & 0.06 & -0.07 & -0.09 & 1 & 0.00 \\ 
0 & 0.05 & 0.39 & 0.07 & 0.11 & 0.01 & 0.00 & 1 \\ 
\end{matrix}
\right).
\end{aligned}
\end{equation}

To help to identify a bi-factor structure from $\hat{\Lambda}^{oblq}$, all loadings whose absolute value is less than 0.2 are set to zero, as is done in \cite{jennrich2012exploratory}. The adjusted loadings are presented in Table~\ref{tab:bfrt-loading-r}. As expected, the loading structure does not conform strictly to a bi-factor model, with four items loading onto three factors.

\begin{table}[h]
    \centering
    \caption{Estimated bi-factor loading matrix with seven group factors. }
    \label{tab:bfrt-loading-r}
    \begin{tabular}{| r | r |r | r | r | r | r | r | r | r |}
    \hline
    Items & Sign & General & G1 & G2 & G3 & G4
    & G5 & G6 & G7\\
    \hline
        1& +E1&0.86 & 0 & 0 & 0 & 0 & 0 & 0.42 & 0 \\ 
        2& +E1 &0.85 & 0 & 0 & 0 & 0 & 0 & 0 & 0 \\ 
        3& $-$E1&0.91 & 0 & 0 & 0 & 0 & 0 & 0 & 0 \\ 
        4& $-$E1&0.87 & 0 & 0 & 0 & 0 & 0 & 0 & -0.20 \\         
        5& +E2&0.88 & 0.69 & 0 & 0 & 0 & 0 & 0 & 0 \\ 
        6& +E2&0.92 & 0.25 & 0 & 0 & 0 & 0 & 0.22 & 0 \\ 
        7& $-$E2&0.72 & 0 & 0 & 0 & 0 & 0 & -0.21 & 0 \\ 
        8& $-$E2&0.85 & 0.22 & 0 & 0 & 0 & 0 & -0.29 & 0 \\ 
        9& +E3&0.52 & 0 & 0 & 0 & 0.79 & 0 & 0 & 0 \\ 
        10& +E3&0.52 & 0 & 0 & 0 & 0.75 & 0 & 0 & 0 \\ 
        11& +E3& 0.44 & 0 & 0 & 0 & 0.62 & 0 & 0 & 0 \\
        12& $-$E3&0.55 & 0 & 0 & 0 & 0.62 & 0 & 0 & 0 \\ 
        13& +E4& 0.32 & 0 & 0 & 0 & 0 & 0.82 & 0 & 0  \\ 
        14& +E4& 0.51 & 0 & 0 & 0 & 0 & 0.74 & 0 & 0 \\ 
        15& +E4&0.49 & 0 & 0 & 0 & 0 & 0.51 & 0 & 0\\ 
        16& $-$E4&0 & 0 & 0 & 0 & 0 & 0.37 & 0 & 0 \\
        17& +E5&0.46 & 0 & 0 & 0 & 0 & 0& 0 & 0.49 \\ 
        18& +E5&0.53 & 0 & 0 & 0 & 0 & 0 & 0 & 0.62 \\ 
        19& +E5&0.28 & 0 & 0.48 & 0 & 0 &0 & 0 & 0.33 \\ 
        20& +E5&0.48 &0 & 1.10 & 0 & 0 & 0 & 0 & 0 \\ 
        21& +E6 & 0.64 & 0 & 0 & 0.26 & 0 & 0 & 0.28 & 0  \\ 
        22& +E6&0.69 & 0 & 0 & 0.40 & 0 & 0 & 0 & 0 \\ 
        23& +E6&0.63 & 0 & 0 & 0.63 & 0 & 0 & 0 & 0 \\ 
        24& +E6&0.58 & 0 & 0 & 0.60 & 0 & 0 & 0 & 0 \\ \hline
    \end{tabular}
\end{table}

We now analyze the estimated model in detail. In this
result, we have adjusted the sign flip and column swapping to align with the result of
the proposed method. All loadings on the general factor are positive, supporting the existence of a general extraversion factor. We interpret the group factors G3, G4, G5 as the Cheerfulness, Assertiveness and Activity Level factors respectively. G2, loaded with the items "19 Enjoy being reckless" and "20 Act wild and crazy" is interpreted as the Reckless Excitement-Seeking factor and consistent with the result from our proposed method. G7 is loaded with items "4 Keep others at a distance", "17 Love excitement", "18 Seek adventure" and "19 Enjoy being reckless". Even though G2 and G7 are loaded with item 19 in common, G7 emphasizes more on the pursuit of meaningful
experiences. So we still interpret G7 as the Meaningful Excitement-Seeking factor. Additionally, G2 and G7 are positively correlated, as is the case in Section~\ref{sec:real data}.

There is a notable difference between the results from the two methods. The result of the ALM method shows the clear presence of a Friendliness factor (G1) and a Gregariousness factor (G6). However, for the bi-factor rotation method, these does not seem to exist a clear Friendliness factor. Both G1 and G6 in the solution of the bi-factor rotation method are related to Gregariousness. Large loadings of the variables designed to measure Friendliness now spread out among several group factors. 

Overall, both methods suggest similar (approximate) bi-factor model structures, and the result from the proposed method tends to be neater and more interpretable.

\section{Technical Proofs}\label{appen:Proof}
\subsection{Proof of Theorem \ref{thm:id}}\label{appen:Proofthm}
Suppose that $\Lambda\Phi(\Lambda)^{\top}+\Psi = \Lambda^{*}\Phi^{*}(\Lambda^{*})^{\top} + \Psi^{*}$. Under Condition \ref{cond:seperate}, we have $\Lambda\Phi(\Lambda)^{\top} = \Lambda^{*}\Phi^{*}(\Lambda^{*})^{\top}$. For the simplicity of the notation, we substitute $\Lambda[\mathcal{B}^{*}_{g},\{1,\ldots,G+1\}]$ for $\Lambda[\mathcal{B}^{*}_{g},:]$. The proof consists of three parts:(1) show the bi-factor structure of $\mathcal{B}_{g_{1}}^{*}$ is unique, (2) show that combined with some group $g_{2}\in\mathcal{H}^{*}, g_{2}\neq g_{1}$, $\Lambda[\mathcal{B}^{*}_{g_{1}},:]$ is identified up to a sign flip and a group permutation, (3) complete the proof of Theorem \ref{thm:id}.

We first consider the equation
\begin{equation}\label{eq: app thm1 eq1}
\Lambda[\mathcal{B}^{*}_{g_1},:]\Phi(\Lambda[\mathcal{B}^{*}_{g_1},:])^{\top} = \Lambda^{*}[\mathcal{B}^{*}_{g_1},\{1,1+g_1\}](\Lambda^{*}[\mathcal{B}^{*}_{g_1},\{1,1+g_1\}])^{\top}.
\end{equation}
Since the matrix on the right side of \eqref{eq: app thm1 eq1} has rank 2, there exist 2 possible bi-factor structures for the matrix on the left side of \eqref{eq: app thm1 eq1}: (1)$\Lambda[\mathcal{B}^{*}_{g_1},\{1,1+g'_{1}\}](\Lambda[\mathcal{B}^{*}_{g_1},\{1,1+g'_{1}\}])^{\top} = \Lambda^{*}[\mathcal{B}^{*}_{g_1},\{1,1+g_1\}](\Lambda^{*}[\mathcal{B}^{*}_{g_1},\{1,1+g_1\}])^{\top}$ for some $g'_{1}\in\{1,\ldots,G\}$ and (2) There exists a partition of $\mathcal{B}^{*}_{g_1} = \mathcal{B}^{*}_{g_{1,1}}\cup \mathcal{B}^{*}_{g_{1,2}}$ and $g'_{1},g'_{2}\in\{1,\ldots,G\}$ such that
\begin{equation}\label{eq: app thm1 eq2}
\begin{aligned}
\begin{pmatrix}
\boldsymbol{\lambda}_{1} & \boldsymbol{\lambda}_{g'_{1}}&\mathbf{0}\\
\boldsymbol{\lambda}_{2} & \mathbf{0} & \boldsymbol{\lambda}_{g'_{2}}
\end{pmatrix}
\begin{pmatrix}
1&0&0\\
0&1&\phi_{1+g'_{1},1+g'_{2}}\\
0&\phi_{1+g'_{1},1+g'_{2}}&1
\end{pmatrix}
\begin{pmatrix}
\boldsymbol{\lambda}_{1}^{\top} & \boldsymbol{\lambda}_{2}^{\top}\\
\boldsymbol{\lambda}_{g'_{1}}^{\top}&\mathbf{0}^{\top}\\
\mathbf{0}^{\top} & \boldsymbol{\lambda}_{g'_{2}}^{\top}
\end{pmatrix}
=\begin{pmatrix}
\boldsymbol{\lambda}^{*}_{1} & \boldsymbol{\lambda}^{*}_{g_{1}}\\
\boldsymbol{\lambda}^{*}_{2} & \boldsymbol{\lambda}^{*}_{g_{2}}
\end{pmatrix}
\begin{pmatrix}
(\boldsymbol{\lambda}^{*}_{1})^{\top} & (\boldsymbol{\lambda}^{*}_{2})^{\top} \\

(\boldsymbol{\lambda}^{*}_{g_{1}})^{\top} & (\boldsymbol{\lambda}^{*}_{g_{2}})^{\top}
\end{pmatrix},
\end{aligned}
\end{equation}
where $\boldsymbol{\lambda}_{i} = \Lambda[\mathcal{B}^{*}_{g_{1,i}},\{1\}]$, $\boldsymbol{\lambda}_{g'_{i}} = \Lambda[\mathcal{B}^{*}_{g_{1,i}},\{1+g'_{i}\}]$, $\boldsymbol{\lambda}^{*}_{i} = \Lambda^{*}[\mathcal{B}^{*}_{g_{1,i}},\{1\}]$ and $\boldsymbol{\lambda}^{*}_{g_{i}} = \Lambda^{*}[\mathcal{B}^{*}_{g_{1,i}},\{1+g_{1}\}]$ with $\boldsymbol{\lambda}_{g'_{i}} \neq \mathbf{0}$ for $i=1,2$.

Here we consider the second case. Since the matrix on the right side of \eqref{eq: app thm1 eq2} has rank 2, we must have $(\boldsymbol{\lambda}_{i},\boldsymbol{\lambda}_{g'_{i}})$ has rank 1 for $i=1,2$, which leads to the fact that $(\boldsymbol{\lambda}^{*}_{i},\boldsymbol{\lambda}^{*}_{g_{i}})$ has rank 1. However, by Condition~\ref{cond:E3Sfang}, there exists at least one of $(\boldsymbol{\lambda}^{*}_{1},\boldsymbol{\lambda}^{*}_{g_{1}})$ and $(\boldsymbol{\lambda}^{*}_{2},\boldsymbol{\lambda}^{*}_{g_{2}})$ has rank 2. Thus, we must have $\Lambda[\mathcal{B}^{*}_{g_1},\{1,1+g'_{1}\}](\Lambda[\mathcal{B}^{*}_{g_1},\{1,1+g'_{1}\}])^{\top} = \Lambda^{*}[\mathcal{B}^{*}_{g_1},\{1,1+g_1\}](\Lambda^{*}[\mathcal{B}^{*}_{g_1},\{1,1+g_1\}])^{\top}$ for some $g'_{1}\in\{1,\ldots,G\}$. Without loss of generation, we assume $g'_1 = g_1$.

Secondly, there exits some $g_{2}\in\mathcal{H}^{*}$ and $g_{2} \neq g_{1}$ by Condition~\ref{cond:E3Sfang}. We consider the $\mathcal{B}^{*}_{g_1}\cup \mathcal{B}^{*}_{g_2}$ rows and $\mathcal{B}^{*}_{g_1}\cup \mathcal{B}^{*}_{g_2}$ columns of $\Lambda\Phi(\Lambda)^{\top}$ and $\Lambda^{*}\Phi^{*}(\Lambda^{*})^{\top}$. Since the bi-factor structure of the $\mathcal{B}^{*}_{g_1}$ rows and $\mathcal{B}^{*}_{g_1}$ columns has already been known, there are two possible bi-factor structures: (1) There exists some $g'_2\in\{1,\ldots, G\}$ such that
\begin{equation}\label{eq: app thm1 eq3}
\begin{aligned}
\begin{pmatrix}
\boldsymbol{\lambda}_{1} & \boldsymbol{\lambda}_{g_{1}}&\mathbf{0}\\
\boldsymbol{\lambda}_{2} & \mathbf{0} & \boldsymbol{\lambda}_{g'_{2}}
\end{pmatrix}
\begin{pmatrix}
1&0&0\\
0&1&\rho_{1,2}\\
0&\rho_{1,2}&1
\end{pmatrix}
\begin{pmatrix}
\boldsymbol{\lambda}_{1}^{\top} & \boldsymbol{\lambda}_{2}^{\top}\\
\boldsymbol{\lambda}_{g_{1}}^{\top}&\mathbf{0}^{\top}\\
\mathbf{0}^{\top} & \boldsymbol{\lambda}_{g'_{2}}^{\top}
\end{pmatrix}
=\begin{pmatrix}
\boldsymbol{\lambda}^{*}_{1} & \boldsymbol{\lambda}^{*}_{g_{1}}&\mathbf{0}\\
\boldsymbol{\lambda}^{*}_{2} & \mathbf{0} & \boldsymbol{\lambda}^{*}_{g_{2}}
\end{pmatrix}
\begin{pmatrix}
1&0&0\\
0&1&\rho_{1,2}^{*}\\
0&\rho_{1,2}^{*}&1
\end{pmatrix}
\begin{pmatrix}
(\boldsymbol{\lambda}_{1}^{*})^{\top} & (\boldsymbol{\lambda}_{2}
^{*})^{\top}\\
(\boldsymbol{\lambda}_{g_{1}}^{*})^{\top}&\mathbf{0}^{\top}\\
\mathbf{0}^{\top} & (\boldsymbol{\lambda}_{g_{2}}^{*})^{\top}
\end{pmatrix},
\end{aligned}
\end{equation}
where $\boldsymbol{\lambda}_{i} = \Lambda[\mathcal{B}^{*}_{g_{i}},\{1\}]$,  $\boldsymbol{\lambda}_{g'_{i}} = \Lambda[\mathcal{B}^{*}_{g_{i}},\{1+g'_{i}\}]$,  $\boldsymbol{\lambda}^{*}_{i} = \Lambda^{*}[\mathcal{B}^{*}_{g_{i}},\{1\}]$ and $\boldsymbol{\lambda}^{*}_{g_{i}} = \Lambda^{*}[\mathcal{B}^{*}_{g_{i}},\{1+g_{1}\}]$ for $i=1,2$. $\rho_{1,2} = \phi_{1+g'_{1},1+g'_{2}}$ and $\rho^{*}_{1,2} = \phi^{*}_{1+g_{1},1+g_{2}}$. 

(2) There exists a partition of $\mathcal{B}^{*}_{g_2} = \mathcal{B}^{*}_{g_{2,1}}\cup \mathcal{B}^{*}_{g_{2,2}}$ and $g'_{2}\in\{1,\ldots,G\}$ such that

\begin{equation}\label{eq: app thm1 eq4}
\begin{aligned}
&\begin{pmatrix}
\boldsymbol{\lambda}_{1} & \boldsymbol{\lambda}_{g_{1}}&\mathbf{0}\\
\boldsymbol{\lambda}_{2,1} & \boldsymbol{\lambda}_{2,g_{1}}&\mathbf{0}\\
\boldsymbol{\lambda}_{2,2} & \mathbf{0} & \boldsymbol{\lambda}_{2,g'_{2}}
\end{pmatrix}
\begin{pmatrix}
1&0&0\\
0&1&\rho_{1,2}\\
0&\rho_{1,2}&1
\end{pmatrix}
\begin{pmatrix}
\boldsymbol{\lambda}_{1}^{\top} & \boldsymbol{\lambda}_{2,1}^{\top} & \boldsymbol{\lambda}_{2,2}^{\top}\\
\boldsymbol{\lambda}_{g_{1}}^{\top}& \boldsymbol{\lambda}_{2,g_{1}}^{\top}&\mathbf{0}^{\top}\\
\mathbf{0}^{\top} & \mathbf{0}^{\top} &\boldsymbol{\lambda}_{2,g'_{2}}^{\top}
\end{pmatrix}\\
=&\begin{pmatrix}
\boldsymbol{\lambda}^{*}_{1} & \boldsymbol{\lambda}^{*}_{g_{1}}&\mathbf{0}\\
\boldsymbol{\lambda}^{*}_{2,1} & \mathbf{0} & \boldsymbol{\lambda}^{*}_{g_{2},1}\\
\boldsymbol{\lambda}^{*}_{2,2} & \mathbf{0} & \boldsymbol{\lambda}^{*}_{g_{2},2}\\
\end{pmatrix}
\begin{pmatrix}
1&0&0\\
0&1&\rho_{1,2}^{*}\\
0&\rho_{1,2}^{*}&1
\end{pmatrix}
\begin{pmatrix}
(\boldsymbol{\lambda}_{1}^{*})^{\top} & (\boldsymbol{\lambda}_{2,1}
^{*})^{\top} & (\boldsymbol{\lambda}_{2,2}
^{*})^{\top}\\
(\boldsymbol{\lambda}_{g_{1}}^{*})^{\top}&\mathbf{0}^{\top}&\mathbf{0}^{\top}\\
\mathbf{0}^{\top} & (\boldsymbol{\lambda}_{g_{2},1}^{*})^{\top}  & (\boldsymbol{\lambda}_{g_{2},2}^{*})^{\top}
\end{pmatrix},
\end{aligned}
\end{equation}
where $\boldsymbol{\lambda}_{1} = \Lambda[\mathcal{B}^{*}_{g_{1}},\{1\}]$, $\boldsymbol{\lambda}_{g_{1}} = \Lambda[\mathcal{B}^{*}_{g_{1}},\{1+g_{1}\}]$, $\boldsymbol{\lambda}^{*}_{1} = \Lambda^{*}[\mathcal{B}^{*}_{g_{1}},\{1\}]$,  $\boldsymbol{\lambda}^{*}_{g_{1}} = \Lambda^{*}[\mathcal{B}^{*}_{g_{1}},\{1+g_{1}\}]$, 
$\boldsymbol{\lambda}_{2,i} = \Lambda[\mathcal{B}^{*}_{g_{2},i},\{1\}]$, 
$\boldsymbol{\lambda}^{*}_{2,i} = \Lambda[\mathcal{B}^{*}_{g_{2},i},\{1\}]$, $\boldsymbol{\lambda}^{*}_{g_{2},i} = \Lambda[\mathcal{B}^{*}_{g_{2},i},\{1+g_2\}]$ for $i=1,2$, $\boldsymbol{\lambda}_{2,g_{1}} = \Lambda[\mathcal{B}^{*}_{g_{2},1},\{1+g_{1}\}]$, $\boldsymbol{\lambda}_{2,g'_{2}} = \Lambda[\mathcal{B}^{*}_{g_{2},2},\{1+g'_{2}\}]$, $\rho_{1,2} = \phi_{1+g_{1},1+g'_{2}}$ and $\rho^{*}_{1,2} = \phi^{*}_{1+g_{1},1+g_{2}}$.

For the second case in \eqref{eq: app thm1 eq4}, there exists some $\alpha$ such that $\boldsymbol{\lambda}_{1} = \cos{\alpha}\boldsymbol{\lambda}^{*}_{1} - \sin{\alpha}\boldsymbol{\lambda}^{*}_{g_{1}}$ and $\boldsymbol{\lambda}_{g_{1}} = \sin{\alpha}\boldsymbol{\lambda}^{*}_{1} + \cos{\alpha}\boldsymbol{\lambda}^{*}_{g_{1}}$. Since the $\mathcal{B}_{g_{2}}^{*}$ rows and the $\mathcal{B}_{g_{2}}^{*}$ columns of $\Lambda\Phi(\Lambda)^{\top}$ and $\Lambda^{*}\Phi^{*}(\Lambda^{*})^{\top}$ have rank 2, under the bi-factor structure of the second case, we have that $(\boldsymbol{\lambda}_{2,1}, \boldsymbol{\lambda}_{2,g_{1}}, \boldsymbol{\lambda}^{*}_{2,1}, \boldsymbol{\lambda}^{*}_{g_{2},1})$ has rank 1 and $(\boldsymbol{\lambda}_{2,2}, \boldsymbol{\lambda}_{2,g'_{2}}, \boldsymbol{\lambda}^{*}_{2,2}, \boldsymbol{\lambda}^{*}_{g_{2},2})$ has rank 1. Noticing that $\boldsymbol{\lambda}^{*}_{g_{2},1} \neq\mathbf{0}$, we assume that $\lambda^{*}_{2,1} = k_{1}\boldsymbol{\lambda}^{*}_{g_{2},1}$, $\boldsymbol{\lambda}_{2,1} = k_{2}\boldsymbol{\lambda}^{*}_{g_{2},1}$ and $\boldsymbol{\lambda}_{2,g_{1}} = k_{3}\boldsymbol{\lambda}^{*}_{g_{2},1}$. For the $\mathcal{B}^{*}_{g_{2},1}$ rows and the $\mathcal{B}^{*}_{g_{2},1}$ columns of \eqref{eq: app thm1 eq4}, we have $1+k_{1}^{2} = k_{2}^{2}+k_{3}^{2}$. For the $\mathcal{B}^{*}_{g_{1}}$ rows and the $\mathcal{B}^{*}_{g_{2},1}$ columns of \eqref{eq: app thm1 eq4}, we have 
\begin{equation}
\begin{aligned}
&\boldsymbol{\lambda}^{*}_{1}(\boldsymbol{\lambda}^{*}_{2,1})^{\top} + \rho^{*}_{1,2}\boldsymbol{\lambda}^{*}_{g_{1}}(\boldsymbol{\lambda}^{*}_{g_{2},1})^{\top} \\
=& \boldsymbol{\lambda}_{1}(\boldsymbol{\lambda}_{2,1})^{\top} + \boldsymbol{\lambda}_{g_{1}}(\boldsymbol{\lambda}_{2,g_{1}})^{\top}\\
=&k_{2}(\cos{\alpha}\boldsymbol{\lambda}^{*}_{1} - \sin{\alpha}\boldsymbol{\lambda}^{*}_{g_{1}})(\boldsymbol{\lambda}^{*}_{g_{2},1})^{\top} + k_{3}(\sin{\alpha}\boldsymbol{\lambda}^{*}_{1} + \cos{\alpha}\boldsymbol{\lambda}^{*}_{g_{1}})(\boldsymbol{\lambda}^{*}_{g_{2},1})^{\top}\\
=&(k_{2}\cos{\alpha} +  k_{3}\sin{\alpha})\boldsymbol{\lambda}^{*}_{1}(\boldsymbol{\lambda}^{*}_{g_{2},1})^{\top} + (k_{3}\cos{\alpha} - k_{2}\sin{\alpha})\boldsymbol{\lambda}^{*}_{g_{1}}(\boldsymbol{\lambda}^{*}_{g_{2},1})^{\top}.
\end{aligned}
\end{equation}
Then, we have $k_{1} = k_{2}\cos{\alpha} +  k_{3}\sin{\alpha}$ and $\rho^{*}_{1,2} = k_{3}\cos{\alpha} - k_{2}\sin{\alpha}$, which leads to $k_{1}^{2} + (\rho^{*}_{1,2})^{2} = k_{2}^{2} + k_{3}^{2}$. Combined with $1+k_{1}^{2} = k_{2}^{2}+k_{3}^{2}$, we have $|\rho^{*}_{1,2}| = 1$, which contradicts to the fact that $\Phi^{*}$ is positive definite. Thus, only the first case is allowed. Without loss of generation, we assume $g'_2 = g_2$. 

For the first case in \eqref{eq: app thm1 eq3}, there exists some $\alpha,\beta$ such that $\boldsymbol{\lambda}_{1} = \cos{\alpha}\boldsymbol{\lambda}^{*}_{1} - \sin{\alpha}\boldsymbol{\lambda}^{*}_{g_{1}}$, $\boldsymbol{\lambda}_{g_{1}} = \sin{\alpha}\boldsymbol{\lambda}^{*}_{1} + \cos{\alpha}\boldsymbol{\lambda}^{*}_{g_{1}}$, $\boldsymbol{\lambda}_{2} = \cos{\beta}\boldsymbol{\lambda}^{*}_{2} - \sin{\beta}\boldsymbol{\lambda}^{*}_{g_{2}}$ and  $\boldsymbol{\lambda}_{g_{2}} = \sin{\beta}\boldsymbol{\lambda}^{*}_{2} + \cos{\beta}\boldsymbol{\lambda}^{*}_{g_{2}}$. We then have the following equation
\begin{equation}
\begin{pmatrix}
\cos{\alpha} & \sin{\alpha}\\
-\sin{\alpha} & \cos{\alpha}
\end{pmatrix}
\begin{pmatrix}
1&0\\
0&\rho_{12}
\end{pmatrix}
\begin{pmatrix}
\cos{\beta} & -\sin{\beta}\\
\sin{\beta} & \cos{\beta}
\end{pmatrix}
=\begin{pmatrix}
1&0\\
0&\rho^{*}_{12}
\end{pmatrix},
\end{equation}
which leads to $1 = \cos{\alpha}\cos{\beta} + \rho_{12}\sin{\alpha}\sin{\beta}$. Since $|\rho_{12}|<1$, we have $\cos{\alpha}\cos{\beta} = 1$ and $\sin{\alpha}\sin{\beta}=0$. Without loss of generation, we assume $\cos{\alpha} = \cos{\beta} =1$. Then we have $\boldsymbol{\lambda}_{1} = \boldsymbol{\lambda}^{*}_{1}$, $\boldsymbol{\lambda}_{g_{1}} = \boldsymbol{\lambda}^{*}_{g_{1}}$, $\boldsymbol{\lambda}_{2} = \boldsymbol{\lambda}^{*}_{2}$, $\boldsymbol{\lambda}_{g_{2}} = \boldsymbol{\lambda}^{*}_{g_{2}}$ and $\rho_{12} = \rho_{12}^{*}$.

For any group $g_{3}\neq g_{1},g_{2}$, we consider the $\mathcal{B}^{*}_{g1}\cup\mathcal{B}^{*}_{g3}$ rows and $\mathcal{B}^{*}_{g1}\cup\mathcal{B}^{*}_{g3}$ columns of $\Lambda\Phi(\Lambda)^{\top}$ and $\Lambda^{*}\Phi^{*}(\Lambda^{*})^{\top}$. Similar to the the proof of $g_{2}$, there exists only one possible bi-factor structure : For some $g'_{3}\in\{1,\ldots,G\}$, we have
\begin{equation}
\begin{aligned}
\begin{pmatrix}
\boldsymbol{\lambda}_{1} & \boldsymbol{\lambda}_{g'_{1}}&\mathbf{0}\\
\boldsymbol{\lambda}_{3} & \mathbf{0} & \boldsymbol{\lambda}_{g'_{3}}
\end{pmatrix}
\begin{pmatrix}
1&0&0\\
0&1&\rho_{1,3}\\
0&\rho_{1,3}&1
\end{pmatrix}
\begin{pmatrix}
\boldsymbol{\lambda}_{1}^{\top} & \boldsymbol{\lambda}_{3}^{\top}\\
\boldsymbol{\lambda}_{g'_{1}}^{\top}&\mathbf{0}^{\top}\\
\mathbf{0}^{\top} & \boldsymbol{\lambda}_{g'_{3}}^{\top}
\end{pmatrix}
=\begin{pmatrix}
\boldsymbol{\lambda}^{*}_{1} & \boldsymbol{\lambda}^{*}_{g_{1}}&\mathbf{0}\\
\boldsymbol{\lambda}^{*}_{3} & \mathbf{0} & \boldsymbol{\lambda}^{*}_{g_{3}}
\end{pmatrix}
\begin{pmatrix}
1&0&0\\
0&1&\rho_{1,3}^{*}\\
0&\rho_{1,3}^{*}&1
\end{pmatrix}
\begin{pmatrix}
(\boldsymbol{\lambda}_{1}^{*})^{\top} & (\boldsymbol{\lambda}_{3}
^{*})^{\top}\\
(\boldsymbol{\lambda}_{g_{1}}^{*})^{\top}&\mathbf{0}^{\top}\\
\mathbf{0}^{\top} & (\boldsymbol{\lambda}_{g_{3}}^{*})^{\top},
\end{pmatrix}
\end{aligned}
\end{equation}
where $\boldsymbol{\lambda}_{i} = \Lambda[\mathcal{B}^{*}_{g_{i}},\{1\}]$,  $\boldsymbol{\lambda}_{g'_{i}} = \Lambda[\mathcal{B}^{*}_{g_{i}},\{1+g'_{i}\}]$,  $\boldsymbol{\lambda}^{*}_{i} = \Lambda^{*}[\mathcal{B}^{*}_{g_{i}},\{1\}]$ and $\boldsymbol{\lambda}^{*}_{g_{i}} = \Lambda^{*}[\mathcal{B}^{*}_{g_{i}},\{1+g_{1}\}]$ for $i=1,3$. $\rho_{1,3} = \phi_{1+g'_{1},1+g'_{3}}$ and $\rho^{*}_{1,3} = \phi^{*}_{1+g_{1},1+g_{3}}$. We then have 
\begin{equation}
\begin{aligned}
&\boldsymbol{\lambda}^{*}_{1}(\boldsymbol{\lambda}^{*}_{3})^{\top} + \rho^{*}_{1,3}\boldsymbol{\lambda}^{*}_{g_{1}}(\boldsymbol{\lambda}^{*}_{g_{3}})^{\top} = \boldsymbol{\lambda}_{1}(\boldsymbol{\lambda}_{3})^{\top} + \rho_{1,3}\boldsymbol{\lambda}_{g_{1}}(\boldsymbol{\lambda}_{g'_{3}})^{\top}\\
&\boldsymbol{\lambda}^{*}_{3}(\boldsymbol{\lambda}^{*}_{3})^{\top} + \boldsymbol{\lambda}^{*}_{g_{3}}(\boldsymbol{\lambda}^{*}_{g_{3}})^{\top} = \boldsymbol{\lambda}_{3}(\boldsymbol{\lambda}_{3})^{\top} + \boldsymbol{\lambda}_{g_{3}}(\boldsymbol{\lambda}_{g_{3}})^{\top}.
\end{aligned}
\end{equation}
Since we have proved $\boldsymbol{\lambda}^{*}_{1} = \boldsymbol{\lambda}_{1}$ and $\boldsymbol{\lambda}^{*}_{g_{1}} = \boldsymbol{\lambda}_{g_{1}}$, we then have $\boldsymbol{\lambda}^{*}_{3} = \boldsymbol{\lambda}_{3}$, $\boldsymbol{\lambda}^{*}_{g_3} = \boldsymbol{\lambda}_{g'_3}$, $\rho^{*}_{1,3} = \rho_{1,3}$ or $\boldsymbol{\lambda}^{*}_{g_3} = -\boldsymbol{\lambda}_{g'_3}$, $\rho^{*}_{1,3} = -\rho_{1,3}$. 

Now, since $\Lambda$ has $G$ group factors, according to the previous proof, each variable belonging to $\mathcal{B}_{i}^{*}$ loads on a unique group factor according to $\Lambda$ and the loadings of the general factor and the group factors are determined up to a sign flip for $i=1,\ldots, G$. Thus, there exist a diagonal sign-flip matrix $D\in \mathcal{D}$ and a permutation matrix $P\in\mathcal{P}$ such that $\Lambda = \Lambda^{*}PD$. It is straightforward to further check that $\Phi = DP^{\top}\Phi^{*}PD $. Thus, the proof is completed.

\subsection{Identifiability of Estimated Bi-factor Structure in Real Data Example}\label{appen:Proofrd}
{For any matrix $A$, we use $\text{rank}(A)$ to denote the rank of $A$.} The following condition is a necessary condition for the identifiability of the extended bi-factor model under a known bi-factor structure, as proposed in Theorem 3 of \cite{fang2021identifiability}.

\begin{condition}\label{cond:nece}
{$\vert \mathcal{B}_{g} \vert \geq 2$ for all $g = 1,\ldots,G$. }
\end{condition}

We then propose the following condition for the identifiability of parameters when the true bi-factor structure is the same as the estimated structure in Section~\ref{sec:real data}.

{
\begin{condition}\label{cond:nodegen}
For any $m\times n$ dimensional submatrix of $\Phi[\{2,\ldots,1+G\},\{2,\ldots,1+G\}]$, $1\leq m,n\leq G$, it's rank is $\min(m,n)$.
\end{condition}
}
\begin{condition}\label{cond:real sepe}
{For any g such that $|\mathcal{B}^{*}_g|\geq 3$, any 2 rows of $\Lambda^{*}[\mathcal{B}_{g}^{*},\{1,1+g\}]$ are linearly independent.}
\end{condition}

\begin{remark}
{Condition~\ref{cond:nodegen} restricts that the correlation matrix of group factors does not degenerate. In Theorem \ref{thm: rd}, we restrict the parametric space of $\Phi$ to the space satisfying Condition~\ref{cond:nodegen}. We note that $\hat{\Lambda}$ in Section~\ref{sec:real data} satisfies Condition~\ref{cond:nodegen}.} Condition~\ref{cond:real sepe} is easy to check in practice and $\hat{\Lambda}$ in Section~\ref{sec:real data} satisfies Condition~\ref{cond:real sepe}.
\end{remark}

\begin{theorem}\label{thm: rd}
Suppose the true bi-factor structure follows $\hat{\Lambda}$ in Section~\ref{sec:real data}. Let $\Lambda^{*}$, $\Phi^{*}$ and $\Psi^{*}$ be the true parameters such that {Conditions~\ref{cond:nece} -\ref{cond:real sepe} are satisfied}. For any parameters $\Lambda$, $\Phi$ and $\Psi$ that satisfy {Conditions~\ref{cond:nece} and \ref{cond:nodegen}} and $\Lambda^{*}\Phi^{*}(\Lambda^{*})^{\top} + \Psi^{*} = \Lambda\Phi(\Lambda)^{\top} + \Psi$, there exists a diagonal sign-flip matrix $D\in\mathcal{D}$ and a permutation matrix $P\in\mathcal{P}$ such that $\Lambda = \Lambda^{*}PD$, $\Phi = DP^{\top}\Phi^{*}PD $ and $\Psi^{*} = \Psi$.
\end{theorem}

\textbf{Proof of Theorem \ref{thm: rd} :} Without loss of generation, we assume that $|\mathcal{B}_{1}^{*}| = |\mathcal{B}_{2}^{*}| = 5$, $|\mathcal{B}_{3}^{*}| = |\mathcal{B}_{4}^{*}| = 4$ and $|\mathcal{B}_{5}^{*}| = |\mathcal{B}_{6}^{*}| = |\mathcal{B}_{7}^{*}| = 2$. Suppose that there exists $\Lambda$, $\Phi$ and $\Psi$ and $\Sigma = \Lambda\Phi\Lambda^{\top} + \Psi$ such that $\Sigma = \Sigma^*$.
The proof consists of two parts: (1) Show that $\Lambda[\cup_{i=1}^{4}\mathcal{B}^{*}_{i},:]$ and $\Lambda^{*}[\cup_{i=1}^{4}\mathcal{B}^{*}_{i},:]$ has the same bi-factor structure. Without loss of generality, we further assume that $\Lambda[\mathcal{B}_{i}^{*},\{1+i\}]\neq \mathbf{0}$ for $i=1,\ldots,4$. We show that there exists some $5\times 5$ sign flip matrix $\tilde{D}$ such that $\Lambda[\cup_{i=1}^{4}\mathcal{B}^{*}_{i},\{1,\ldots,5\}] = \Lambda^{*}[\cup_{i=1}^{4}\mathcal{B}^{*}_{i},\{1,\ldots,5\}]\tilde{D}$, $\Phi[\{1,\ldots,5\},\{1,\ldots,5\}] = \tilde{D}\Phi^{*}[\{1,\ldots,5\},\{1,\ldots,5\}]\tilde{D}$ and $\psi_{j} = \psi^{*}_{j}$ for $j\in \cup_{i=1}^{4}\mathcal{B}^{*}_{i}$. (2) Show that $\Lambda$ and $\Lambda^{*}$ have the same bi-factor structure for the rest of the variables and complete the proof. 

We now prove the first part. Let $\mathcal{F}_{i} = \{g: \Lambda[\mathcal{B}_{i}^{*},\{1+g\}]\neq \mathbf{0}\} \cup \{1\}$ be the set of factors such that the variables belonging to 
$\mathcal{B}_{i}^{*}$ load on these factors for $i=1,\ldots,4$. We note that $|\mathcal{F}_{i}|\geq 2$ for $i=1,\ldots,4$. When $|\mathcal{F}_{i}|=2$, all variables that belong to $\mathcal{B}_{i}^{*}$ load on the same group factor. We claim that
\begin{equation}\label{eq:app claim}
\text{rank}(\Lambda[\mathcal{B}^{*}_{i},\mathcal{F}_{i}]) = |\mathcal{F}_{i}| \mbox{~~if~~} |\mathcal{F}_{i}|\leq |\mathcal{B}^{*}_{i}| \mbox{~~for~~} i=1,\ldots,4. 
\end{equation} 
If $|\mathcal{F}_{i}|\leq |\mathcal{B}^{*}_{i}|$, there exists some $g_i \in \mathcal{F}_i$, $g_i\neq 1$ and $j_{g_i}, j'_{g_i} \in\mathcal{B}_{i}^{*}$ such that $\lambda_{j_{g_i},1+g_i}\neq 0 $ and $\lambda_{j'_{g_i},1+g_i}\neq 0$. For $1\leq i' \leq 4$, $i'\neq i$, consider the equation $\Sigma[\{j_{g_{i}}, j'_{g_{i}}\},\mathcal{B}_{i'}^{*}] = \Sigma^{*}[\{j_{g_{i}}, j'_{g_{i}}\},\mathcal{B}_{i'}^{*}]$, which is equivalent to 
\begin{equation}\label{eq:app off diag 1}
\begin{aligned}
&\Lambda[\{j_{g_i},j'_{g_i}\},\{1,1+g_{i}\}]\Phi[\{1,1+g_i\},\mathcal{F}_{i'}](\Lambda[\mathcal{B}_{i'}^{*},\mathcal{F}_{i'}])^{\top}\\
=&\Lambda^{*}[\{j_{g_i},j'_{g_i}\},\{1,1+i\}]\Phi^{*}[\{1,1+i\},\{1,1+i'\}](\Lambda^{*}[\mathcal{B}_{i'}^{*},\{1,1+i'\}])^{\top}.
\end{aligned}
\end{equation}
Noticing that by Condition~\ref{cond:nodegen} and \ref{cond:real sepe} hold for $\Phi^*$ and $\Lambda^*$, $$\Lambda^{*}[\{j_{g_i},j'_{g_i}\},\{1,1+i\}]\Phi^{*}[\{1,1+i\},\{1,1+i'\}](\Lambda^{*}[\mathcal{B}_{i'}^{*},\{1,1+i'\}])^{\top}$$ has rank 2. Thus, $\Lambda[\{j_{g_i},j'_{g_i}\},\{1,1+g_{i}\}]$ should have rank 2. Otherwise, $\Lambda[\{j_i,j_{i}'\},\{1,1+g_{i}\}]\Phi[\{1,1+g_i\},\mathcal{F}_{i'}](\Lambda[\mathcal{B}_{i'}^{*},\mathcal{F}_{i'}])^{\top}$ has at most rank 1, which contradicts \eqref{eq:app off diag 1}. Then, since for each $g'_{i}\in\mathcal{F}_i$, $g'_{i}\neq 1$, there exists some $j_{g'_i}$ such that $\lambda_{j_{g'_i},g'_{i}} \neq 0$, it is easy to check that \eqref{eq:app claim} holds.

Then, consider the equation $\Sigma[\mathcal{B}_{i}^{*},\mathcal{B}_{i'}^{*}] = \Sigma^{*}[\mathcal{B}_{i}^{*},\mathcal{B}_{i'}^{*}]$ for $1\leq i\neq i'\leq 4$, which is equivalent to
\begin{equation}\label{eq:app off diag 2}
\Lambda[\mathcal{B}_{i}^{*},\mathcal{F}_{i}]\Phi[\mathcal{F}_{i},\mathcal{F}_{i'}](\Lambda[\mathcal{B}_{i'}^{*},\mathcal{F}_{i'}])^{\top} = \Lambda^{*}[\mathcal{B}_{i}^{*},\{1,1+i\}]\Phi[\{1,1+i\},\{1,1+i'\}](\Lambda[\mathcal{B}_{i'}^{*},\{1,1+i'\}])^{\top}.
\end{equation}
With the same argument, $\Lambda^{*}[\mathcal{B}_{i}^{*},\{1,1+i\}]\Phi[\{1,1+i\},\{1,1+i'\}](\Lambda[\mathcal{B}_{i'}^{*},\{1,1+i'\}])^{\top}$ has rank 2. By Sylvester's rank inequality, we have 
\begin{equation}\label{eq:app rank sy}
\begin{aligned}
&\text{rank}(\Lambda[\mathcal{B}_{i}^{*},\mathcal{F}_{i}]\Phi[\mathcal{F}_{i},\mathcal{F}_{i'}](\Lambda[\mathcal{B}_{i'}^{*},\mathcal{F}_{i'}])^{\top})\\
\geq &\text{rank}((\Lambda[\mathcal{B}_{i}^{*},\mathcal{F}_{i}]) +\text{rank}(\Phi[\mathcal{F}_{i},\mathcal{F}_{i'}]) + \text{rank}(\Lambda[\mathcal{B}_{i'}^{*},\mathcal{F}_{i'}]) - |\mathcal{F}_{i}| - |\mathcal{F}_{i'}|.
\end{aligned}
\end{equation}
We consider the following case
\begin{enumerate}
    \item $|\mathcal{F}_i|\leq |\mathcal{B}_{i}^{*}|$ for all $1\leq i\leq 4$. In this case, according to claim \eqref{eq:app claim} and Condition \ref{cond:nodegen}, inequality \eqref{eq:app rank sy} leads to $$\text{rank}(\Lambda[\mathcal{B}_{i}^{*},\mathcal{F}_{i}]\Phi[\mathcal{F}_{i},\mathcal{F}_{i'}](\Lambda[\mathcal{B}_{i'}^{*},\mathcal{F}_{i'}])^{\top}) \geq \min(|\mathcal{F}_i|,|\mathcal{F}_{i'}|),$$
    for any $1\leq i\neq i'\leq 4$. We then have $\min(|\mathcal{F}_i|,|\mathcal{F}_{i'}|) = 2$. By applying this argument to all pairs $(i,i')$, $1\leq i < i' \leq 4$, we conclude that there exists at most one $\mathcal{F}_i$ such that $|\mathcal{F}_i|\geq 3$ for $1\leq i \leq 4$.

    If there exists some $i$ such that $|\mathcal{F}_i| \geq 3$ for $i=1,\ldots,4$. Without loss of generality, we assume $|\mathcal{F}_1| \geq 3$ and $|\mathcal{F}_i| = 2$ for $i=2,3,4$. We claim that $|\mathcal{F}_{2}\cup\mathcal{F}_{3}| = 3$, in other words, the variables belonging to $\mathcal{B}_{2}^{*}$ and $\mathcal{B}_{3}^{*}$ load on different factors. Otherwise, $|\mathcal{F}_{2}\cup\mathcal{F}_{3}| = 2$. Consider the equation $$\Sigma[\mathcal{B}_{2}^{*}\cup\mathcal{B}_{3}^{*},\mathcal{B}_{2}^{*}\cup\mathcal{B}_{3}^{*}] = \Sigma^{*}[\mathcal{B}_{2}^{*}\cup\mathcal{B}_{3}^{*},\mathcal{B}_{2}^{*}\cup\mathcal{B}_{3}^{*}],$$
    which is equivalent to 
    \begin{equation}\label{eq: app diag}
    \begin{aligned}    &\Lambda[\mathcal{B}_{2}^{*}\cup\mathcal{B}_{3}^{*},\mathcal{F}_{2}\cup\mathcal{F}_{3}]\Phi[\mathcal{F}_{2}\cup\mathcal{F}_{3},\mathcal{F}_{2}\cup\mathcal{F}_{3}](\Lambda[\mathcal{B}_{2}^{*}\cup\mathcal{B}_{3}^{*},\mathcal{F}_{2}\cup\mathcal{F}_{3}])^{\top} +\Psi[\mathcal{B}_{2}^{*}\cup\mathcal{B}_{3}^{*},\mathcal{B}_{2}^{*}\cup\mathcal{B}_{3}^{*}] \\
    =& \Lambda^{*}[\mathcal{B}_{2}^{*}\cup\mathcal{B}_{3}^{*},\{1,3,4\}]\Phi^{*}[\{1,3,4\},\{1,3,4\}](\Lambda^{*}[\mathcal{B}_{2}^{*}\cup\mathcal{B}_{3}^{*},\{1,3,4\}])^{\top} + \Psi^{*}[\mathcal{B}_{2}^{*}\cup\mathcal{B}_{3}^{*},\mathcal{B}_{2}^{*}\cup\mathcal{B}_{3}^{*}].
    \end{aligned}
    \end{equation}
    Since $\Lambda^*$ satisfies Condition~\ref{cond:real sepe}, noticing that $|\mathcal{B}_{2}^{*}|\geq 4$ and $|\mathcal{B}_{3}^{*}|\geq 4$, it is easy to check that the matrix $\Lambda^{*}[\mathcal{B}_{2}^{*}\cup\mathcal{B}_{3}^{*},\{1,3,4\}]$ satisfies the condition for Theorem 5.1 of \cite{anderson1956statistical}, that is, if any row of $\Lambda^{*}[\mathcal{B}_{2}^{*}\cup\mathcal{B}_{3}^{*},\{1,3,4\}]$ is deleted, there still remains two disjoint submatrices of $\Lambda^{*}[\mathcal{B}_{2}^{*}\cup\mathcal{B}_{3}^{*},\{1,3,4\}]$ with rank 3. By applying Theorem 5.1 of \cite{anderson1956statistical}, we have $\Psi[\mathcal{B}_{2}^{*}\cup\mathcal{B}_{3}^{*},\mathcal{B}_{2}^{*}\cup\mathcal{B}_{3}^{*}] = \Psi^{*}[\mathcal{B}_{2}^{*}\cup\mathcal{B}_{3}^{*},\mathcal{B}_{2}^{*}\cup\mathcal{B}_{3}^{*}]$. Thus, we further have
    \begin{equation}\label{eq: app diag sep}
    \begin{aligned}    &\Lambda[\mathcal{B}_{2}^{*}\cup\mathcal{B}_{3}^{*},\mathcal{F}_{2}\cup\mathcal{F}_{3}]\Phi[\mathcal{F}_{2}\cup\mathcal{F}_{3},\mathcal{F}_{2}\cup\mathcal{F}_{3}](\Lambda[\mathcal{B}_{2}^{*}\cup\mathcal{B}_{3}^{*},\mathcal{F}_{2}\cup\mathcal{F}_{3}])^{\top}   \\
    =& \Lambda^{*}[\mathcal{B}_{2}^{*}\cup\mathcal{B}_{3}^{*},\{1,3,4\}]\Phi^{*}[\{1,3,4\},\{1,3,4\}](\Lambda^{*}[\mathcal{B}_{2}^{*}\cup\mathcal{B}_{3}^{*},\{1,3,4\}])^{\top} .
    \end{aligned}
    \end{equation}
    If $|\mathcal{F}_{2}\cup\mathcal{F}_{3}| = 2$, then the rank of the matrix in the first line of \eqref{eq: app diag sep} is 2, which contradicts the fact that the rank of the matrix in the second line of \eqref{eq: app diag sep} is 3. Thus, $|\mathcal{F}_{2}\cup\mathcal{F}_{3}| = 3$. We note that with a similar argument used in \eqref{eq: app diag} and \eqref{eq: app diag sep}, we also have $|\mathcal{F}_{2}\cup\mathcal{F}_{4}| = 3$, $|\mathcal{F}_{3}\cup\mathcal{F}_{4}| = 3$ and $|\cup_{i=2,3,4}\mathcal{F}_{i}| = 4$. Then, consider the equation $$\Sigma[\mathcal{B}_{1}^{*},\mathcal{B}_{2}^{*}\cup\mathcal{B}_{3}^{*}] = \Sigma^{*}[\mathcal{B}_{1}^{*},\mathcal{B}_{2}^{*}\cup\mathcal{B}_{3}^{*}],$$
    which is equivalent to 
    \begin{equation}\label{eq: app off diag 1 vs 3}
    \begin{aligned}    &\Lambda[\mathcal{B}_{1}^{*},\mathcal{F}_1]\Phi[\mathcal{F}_1,\mathcal{F}_{2}\cup\mathcal{F}_{3}](\Lambda[\mathcal{B}_{2}^{*}\cup\mathcal{B}_{3}^{*},\mathcal{F}_{2}\cup\mathcal{F}_{3})^{\top} \\
    =& \Lambda^{*}[\mathcal{B}_{1}^{*},\{1,2\}]\Phi^{*}[\{1,2\},\{1,3,4\}](\Lambda^{*}[\mathcal{B}_{2}^{*}\cup\mathcal{B}_{3}^{*},\{1,3,4\}])^{\top}.
    \end{aligned}
    \end{equation}
    We note that the rank of the matrix in the second line of \eqref{eq: app off diag 1 vs 3} is 2. According to Sylvester’s rank inequality
    \begin{equation}
    \begin{aligned}
    &\text{rank}(\Lambda[\mathcal{B}_{1}^{*},\mathcal{F}_1]\Phi[\mathcal{F}_1,\mathcal{F}_{2}\cup\mathcal{F}_{3}](\Lambda[\mathcal{B}_{2}^{*}\cup\mathcal{B}_{3}^{*},\mathcal{F}_{2}\cup\mathcal{F}_{3})^{\top})\\
    \geq& \text{rank}(\Lambda[\mathcal{B}_{1}^{*},\mathcal{F}_1]) +\text{rank}(\Phi[\mathcal{F}_1,\mathcal{F}_{2}\cup\mathcal{F}_{3}]) + \text{rank}(\Lambda[\mathcal{B}_{2}^{*}\cup\mathcal{B}_{3}^{*},\mathcal{F}_{2}\cup\mathcal{F}_{3}]) - |\mathcal{F}_1| -3\\
    =& |\mathcal{F}_1| + \min(|\mathcal{F}_1|,3) + 3- |\mathcal{F}_1| -3\\
    =& 3,
    \end{aligned}
    \end{equation}
    which contradicts \eqref{eq: app off diag 1 vs 3}.
    
    Thus, in the case, $|\mathcal{F}_i| = 2$ for $i=1,\ldots,4$. Consider the equation     
    \begin{equation}\label{eq: app diag 2}    \Sigma[\cup_{i=1,\ldots,4}\mathcal{B}_{i}^{*},\cup_{i=1,\ldots,4}\mathcal{B}_{i}^{*}] = \Sigma^{*}[\cup_{i=1,\ldots,4}\mathcal{B}_{i}^{*},\cup_{i=1,\ldots,4}\mathcal{B}_{i}^{*}].
    \end{equation}
    By the similar argument discussed in \eqref{eq: app diag}, \eqref{eq: app diag sep} and further applying Theorem 1 to \eqref{eq: app diag 2}, we conclude that in this case, $\Lambda[\cup_{i=1}^{4}\mathcal{B}^{*}_{i},:]$ and $\Lambda^{*}[\cup_{i=1}^{4}\mathcal{B}^{*}_{i},:]$ has the same bi-factor structure. Without loss of generality, we further assume that $\mathcal{F}_i = \{1,1+i\}$ for $i=1,\ldots,4$. Then, there exists some $5\times 5$ sign flip matrix $\tilde{D}$ such that $\Lambda[\cup_{i=1}^{4}\mathcal{B}^{*}_{i},\{1,\ldots,5\}] = \Lambda^{*}[\cup_{i=1}^{4}\mathcal{B}^{*}_{i},\{1,\ldots,5\}]\tilde{D}$, $\Phi[\{1,\ldots,5\},\{1,\ldots,5\}] = \tilde{D}\Phi^{*}[\{1,\ldots,5\},\{1,\ldots,5\}]\tilde{D}$ and $\psi_{j} = \psi^{*}_{j}$ for $j\in \cup_{i=1}^{4}\mathcal{B}^{*}_{i}$.

    \item There exists some $1\leq i\leq 4$ such that $|\mathcal{F}_i| = 1+|\mathcal{B}_{i}^{*}|\geq 5$. In this case, according to \eqref{eq:app rank sy} 
    \begin{equation}
    \text{rank}(\Lambda[\mathcal{B}_{i}^{*},\mathcal{F}_{i}]\Phi[\mathcal{F}_{i},\mathcal{F}_{i'}](\Lambda[\mathcal{B}_{i'}^{*},\mathcal{F}_{i'}])^{\top})\geq 3 \mbox{~~if~~} |\mathcal{F}_{i'}|\geq 4.
    \end{equation}
    Thus, $|\mathcal{F}_{i'}|\leq 3 < |\mathcal{B}^{*}_{i'}|$ for all $1\leq i'\leq 4$, $i'\neq i$. Without loss of generality, let $i=1$. For $i'=2,3,4$, by the same argument in case 1, we have $\mathcal{F}_{2} = \{1,1+g_{2}\}$, $\mathcal{F}_{3} = \{1,1+g_{3}\}$ and $\mathcal{F}_{4} = \{1,1+g_{4}\}$ for different $g_2, g_3$ and $g_4$. Moreover, $\text{rank}(\Lambda[\cup_{i=2,3,4}\mathcal{B}_{i}^{*},\cup_{i=2,3,4}\mathcal{F}_{i}]) = 4$.

    Then, consider the equation $\Sigma[\mathcal{B}^{*}_{1},\cup_{i=2,3,4}\mathcal{B}_{i}^{*}] = \Sigma^{*}[\mathcal{B}^{*}_{1},\cup_{i=2,3,4}\mathcal{B}_{i}^{*}]$, which is equivalent to 
    \begin{equation}\label{eq:app sparse case}
    \begin{aligned}    &\Lambda[\mathcal{B}_{1}^{*},\mathcal{F}_1]\Phi[\mathcal{F}_1,\cup_{i=2,3,4}\mathcal{F}_{i}](\Lambda[\cup_{i=2,3,4}\mathcal{B}_{i}^{*},\cup_{i=2,3,4}\mathcal{F}_{i}])^{\top}\\
    =&\Lambda^{*}[\mathcal{B}_{1}^{*},\{1,2\}]\Phi^{*}[\{1,2\},\{1,3,4,5\}](\Lambda[\cup_{i=2,3,4}\mathcal{B}_{i}^{*},\{1,3,4,5\}])^{\top}
    \end{aligned}
    \end{equation}
    It is straightforward that $\Lambda^{*}[\mathcal{B}_{1}^{*},\{1,2\}]\Phi^{*}[\{1,2\},\{1,3,4,5\}](\Lambda[\cup_{i=2,3,4}\mathcal{B}_{i}^{*},\{1,3,4,5\}])^{\top}$ has rank 2 according to Condition~\ref{cond:nodegen} and \ref{cond:real sepe}. While, since $|\cup_{i=2,3,4}\mathcal{F}_{i}| = 4 < |\mathcal{F}_1|$, according to Sylvester's rank inequality, 
    \begin{equation}
    \begin{aligned}
    &\text{rank}(\Lambda[\mathcal{B}_{1}^{*},\mathcal{F}_1]\Phi[\mathcal{F}_1,\cup_{i=2,3,4}\mathcal{F}_{i}](\Lambda[\cup_{i=2,3,4}\mathcal{B}_{i}^{*},\cup_{i=2,3,4}\mathcal{F}_{i}])^{\top})\\
    \geq & \text{rank}(\Lambda[\mathcal{B}_{1}^{*},\mathcal{F}_1]) + \text{rank}(\Phi[\mathcal{F}_1,\cup_{i=2,3,4}\mathcal{F}_{i}]) + \text{rank}(\Lambda[\cup_{i=2,3,4}\mathcal{B}_{i}^{*},\cup_{i=2,3,4}\mathcal{F}_{i}]) - |\mathcal{F}_1| - 4\\
    = & |\mathcal{F}_1|-1 +4 +4 - |\mathcal{F}_1| - 4\\
    = & 3,
    \end{aligned}
    \end{equation}
    which contradicts to equation \eqref{eq:app sparse case}. Thus, this case does not exist.
\end{enumerate}

Next, we prove the second part. We denote by $\mathcal{B}_{5}^{*} = \{j_{5},j_{6}\}$, $\mathcal{B}_{6}^{*} = \{j_{7},j_{8}\}$ and $\mathcal{B}_{7}^{*} = \{j_{9},j_{10}\}$. Since $\Lambda$, $\Phi$ and $\Psi$ satisfy Condition~\ref{cond:nece}, there exists three types of possible of bi-factor structure of $\mathcal{B}_{5}$, $\mathcal{B}_{6}$ and $\mathcal{B}_{7}$ and we discuss the three cases one by one. Without loss of generality, we assume $D'$ given in the first part equals the identity matrix.
\begin{enumerate}
    \item None of the bi-factor structures of the variables belonging to $\mathcal{B}_{i}^{*}$, $i=5,6,7$, is correct. Without loss of generality, we assume $\mathcal{B}_{5} = \{j_{5},j_{10}\}$, $\mathcal{B}_{6} = \{j_{6},j_{7}\}$ and $\mathcal{B}_{7} = \{j_{8},j_{9}\}$. In this case, we consider the equation
    \begin{equation}
    \Sigma[\mathcal{B}_{1}^{*},\mathcal{B}_{5}^{*}] = \Sigma^{*}[\mathcal{B}_{1}^{*},\mathcal{B}_{5}^{*}],
    \end{equation}
    which is equivalent to 
    \begin{equation}
    \begin{aligned}
    &\Lambda[\mathcal{B}^{*}_{1},\{1\}]\lambda_{j_{5},1} + \phi_{2,6}\Lambda[\mathcal{B}^{*}_{1},\{2\}]\lambda_{j_{5},6} = \Lambda^{*}[\mathcal{B}^{*}_{1},\{1\}]\lambda^{*}_{j_{5},1} + \phi^{*}_{2,6}\Lambda^{*}[\mathcal{B}^{*}_{1},\{2\}]\lambda^{*}_{j_{5},6},\\
    &\Lambda[\mathcal{B}^{*}_{1},\{1\}]\lambda_{j_{6},1} + \phi_{2,7}\Lambda[\mathcal{B}^{*}_{1},\{2\}]\lambda_{j_{6},7} = \Lambda^{*}[\mathcal{B}^{*}_{1},\{1\}]\lambda^{*}_{j_{6},1} + \phi^{*}_{2,6}\Lambda^{*}[\mathcal{B}^{*}_{1},\{2\}]\lambda^{*}_{j_{6},6}.
    \end{aligned}
    \end{equation}
    Since the first part is proved, we have assumed that $\Lambda[\mathcal{B}^{*}_{1},\{1\}] = \Lambda^{*}[\mathcal{B}^{*}_{1},\{1\}]$ and $\Lambda[\mathcal{B}^{*}_{1},\{2\}] = \Lambda^{*}[\mathcal{B}^{*}_{1},\{2\}]$. Noticing that $\Lambda^{*}[\mathcal{B}^{*}_{1},\{1\}]$ and $\Lambda^{*}[\mathcal{B}^{*}_{1},\{2\}]$ are linearly independent, we have $\lambda_{j_{5},1} = \lambda^{*}_{j_{5},1}$ and  $\lambda_{j_{6},1} = \lambda^{*}_{j_{6},1}$. Similarly, by considering the equations $\Sigma[\mathcal{B}_{1}^{*},\mathcal{B}_{6}^{*}] = \Sigma^{*}[\mathcal{B}_{1}^{*},\mathcal{B}_{6}^{*}]$ and  $\Sigma[\mathcal{B}_{1}^{*},\mathcal{B}_{7}^{*}] = \Sigma^{*}[\mathcal{B}_{1}^{*},\mathcal{B}_{7}^{*}]$, we have $\lambda_{j_{7},1} = \lambda^{*}_{j_{7},1}$, $\lambda_{j_{8},1} = \lambda^{*}_{j_{8},1}$, $\lambda_{j_{9},1} = \lambda^{*}_{j_{9},1}$ and $\lambda_{j_{10},1} = \lambda^{*}_{j_{10},1}$. 
        
    By considering the equation
    \begin{equation}
    \Sigma[j_5,j_6] = \Sigma^{*}[j_5,j_6],
    \end{equation}
    which is equivalent to $\lambda_{j_{5},1}\lambda_{j_{6},1} + \phi_{6,7}\lambda_{j_{5},6}\lambda_{j_{6},7} = \lambda^{*}_{j_{5},1}\lambda^{*}_{j_{6},1} + \lambda^{*}_{j_{5},6}\lambda^{*}_{j_{6},6}$. We further have
    \begin{equation}\label{eq: group loading cross case1 1}
    \phi_{6,7}\lambda_{j_{5},6}\lambda_{j_{6},7} = \lambda^{*}_{j_{5},6}\lambda^{*}_{j_{6},6}.
    \end{equation}
    We can similarly have the equations 
    \begin{equation}\label{eq: group loading cross case1 2}
    \begin{aligned}
    \phi_{6,7}\lambda_{j_{5},6}\lambda_{j_{7},7} &=& \phi_{6,7}^{*}\lambda^{*}_{j_{5},6}\lambda^{*}_{j_{7},7}, \\\phi_{6,7}\lambda_{j_{10},6}\lambda_{j_{6},7} &=& \phi_{6,8}^{*}\lambda^{*}_{j_{10},8}\lambda^{*}_{j_{6},6}, \\\phi_{6,7}\lambda_{j_{10},6}\lambda_{j_{7},8} &=& \phi_{7,8}^{*}\lambda^{*}_{j_{10},8}\lambda^{*}_{j_{7},7}.
    \end{aligned}
    \end{equation}
    By combining the equations \eqref{eq: group loading cross case1 1} and \eqref{eq: group loading cross case1 2}, we have $\phi^{*}_{6,7}\phi^{*}_{6,8} = \phi^{*}_{7,8}$. In a symmetric manner, we further have $\phi^{*}_{6,8}\phi^{*}_{7,8} = \phi^{*}_{6,7}$ and $\phi^{*}_{6,7}\phi^{*}_{7,8} = \phi^{*}_{6,8}$. Since $\phi^{*}_{6,7}, \phi^{*}_{6,8}$ and $  \phi^{*}_{7,8}\neq 0$, we have $|\phi^{*}_{6,7}\phi^{*}_{6,8}\phi^{*}_{7,8}| =1$, which leads to $|\phi^{*}_{6,7}| = |\phi^{*}_{6,8}| = |\phi^{*}_{7,8}| = 1$ and violates the assumption that $\Phi^{*}$ is positive definite. Thus, this case does not exist.
        
    \item Only one of the bi-factor structures of the variables belonging to $\mathcal{B}_{i}^{*}$, $i=5,6,7$, is correct. Without loss of generality, we assume $\mathcal{B}_{5} = \{j_{5},j_{6}\}$, $\mathcal{B}_{6} = \{j_{7},j_{9}\}$ and $\mathcal{B}_{7} = \{j_{8},j_{10}\}$. By the same argument in the first case, we have $\lambda_{j_{i},1} = \lambda^{*}_{j_{i},1}$ for $i=5,\ldots,10$. Next, consider the equations on the diagonal entries of 
    \begin{equation}
    \Sigma[\mathcal{B}_{6}^{*}\cup \mathcal{B}_{7}^{*},\mathcal{B}_{6}^{*}\cup \mathcal{B}_{7}^{*}] = \Sigma^{*}[\mathcal{B}_{6}^{*}\cup \mathcal{B}_{7}^{*},\mathcal{B}_{6}^{*}\cup \mathcal{B}_{7}^{*}].
    \end{equation}
    we have the following 6 equations
    \begin{equation}
    \begin{aligned}
    &\phi_{7,8}\lambda_{j_{7},7}\lambda_{j_{8},8} = \lambda^{*}_{j_{7},7}\lambda^{*}_{j_{8},7}, \\ 
    &\lambda_{j_{7},7}\lambda_{j_{9},7} = \phi^{*}_{7,8}\lambda^{*}_{j_{7},7}\lambda^{*}_{j_{9},8},\\
    &\phi_{7,8}\lambda_{j_{7},7}\lambda_{j_{10},8} = \phi^{*}_{7,8}\lambda^{*}_{j_{7},7}\lambda^{*}_{j_{10},8},\\
    &\phi_{7,8}\lambda_{j_{8},8}\lambda_{j_{9},7} = \phi^{*}_{7,8}\lambda^{*}_{j_{8},7}\lambda^{*}_{j_{9},8},\\
    &\lambda_{j_{8},8}\lambda_{j_{10},8} = \phi^{*}_{7,8}\lambda^{*}_{j_{8},7}\lambda^{*}_{j_{10},8},\\
    &\phi_{7,8}\lambda_{j_{9},7}\lambda_{j_{10},8} = \lambda^{*}_{j_{9},8}\lambda^{*}_{j_{10},8}.
    \end{aligned}
    \end{equation}
    According to the first equation above, we have $\phi_{7,8}\neq 0$. By the 6 equations, we also have $(\phi^{*}_{7,8})^{4}\phi_{7,8}^{4} = \phi_{7,8}^{2}$. Then we have $|\phi^{*}_{7,8}| = |\phi_{7,8}| = 1$, which violates the assumption that $\Phi^{*}$ is positive definite. Thus, this case does not exist.

    \item The bi-factor structure is correct. Without loss of generality, we assume $\mathcal{B}_{5} = \{j_{5},j_{6}\}$, $\mathcal{B}_{6} = \{j_{7},j_{8}\}$ and $\mathcal{B}_{7} = \{j_{9},j_{10}\}$. Similar to the previous argument, we first have $\lambda_{j_{i},1} = \lambda^{*}_{j_{i},1}$ for $i=5,\ldots,10$. Moreover, we have $\phi_{2,6}\lambda_{5,6} = \phi^{*}_{2,6}\lambda^{*}_{5,6}$, $\phi_{2,6}\lambda_{6,6} = \phi^{*}_{2,6}\lambda^{*}_{6,6}$ and $\lambda_{5,6}\lambda_{6,6} = \lambda^{*}_{5,6}\lambda^{*}_{6,6}$. These 3 equations leads to $\phi_{2,6} = \phi^{*}_{2,6}$, $\lambda_{5,6} = \lambda^{*}_{5,6}$ and $\lambda_{6,6} = \lambda^{*}_{6,6}$ or $\phi_{2,6} = -\phi^{*}_{2,6}$, $\lambda_{5,6} = -\lambda^{*}_{5,6}$ and $\lambda_{6,6} = -\lambda^{*}_{6,6}$. With the same argument, the loadings and correlations related with the variables belonging to $\mathcal{B}_{6}^{*}$ and $\mathcal{B}_{7}^{*}$ are also determined up to a sign flip. The check of $\psi_j = \psi^{*}_j$ for $j\in\mathcal{B}_{i}^{*}$, $i=5,6,7$ are straight forward.
\end{enumerate}

\vspace{1cm}
\bibliographystyle{apalike}
\bibliography{ref}
\end{document}